\documentclass{article}

\usepackage[english]{babel}
\usepackage[dvipsnames]{xcolor}

\newcommand{\nn}{\nonumber \\}

\newcommand{\ma}{\mathrm{a}}
\newcommand{\mb}{\mathrm{b}}
\newcommand{\mc}{\mathrm{c}}

\newcommand{\dd}{\mathop{}\!\mathrm{d}}

\usepackage[letterpaper,top=2cm,bottom=2cm,left=3cm,right=3cm,marginparwidth=1.75cm]{geometry}

\usepackage{amsmath}
\usepackage{amssymb}
\usepackage{graphicx}
\usepackage{mathtools}
\usepackage[colorlinks=true, allcolors=blue]{hyperref}
\hypersetup{linktoc=page}
\makeatletter
\renewcommand{\@dotsep}{10000}
\makeatother

\usepackage[T1]{fontenc} 
\usepackage{currvita}

\newcommand{\pp}{=\kern-0.40em{\vert}}

\date{}

\begin{document}

\begin{center}

\vspace*{0.5cm}

{\LARGE
W-symmetries, anomalies and heterotic backgrounds\\
with $G_2$ holonomy
\par}

\thispagestyle{empty}
\vspace{1.2cm}

{\large Loukas Grimanellis$^{1,2}$}

\vspace{0.7cm}

{\small
$^1$\textit{Division of Theoretical Physics,
Ru\dj er Bo\v{s}kovi\'c Institute,
Bijeni\v{c}ka 54, 10000 Zagreb, Croatia}

\vspace{0.2cm}

$^2$\textit{Department of Mathematics,
King's College London,
Strand, London WC2R 2LS, UK}

\vspace{0.45cm}

\texttt{loukas.grimanellis@irb.hr}
}

\end{center}

\vspace{1.8cm}

\begin{abstract}
\small
\noindent
Heterotic sigma models on backgrounds with $G_2$ holonomy admit a class of symmetries generated by forms that are covariantly constant with respect to a connection with skew-symmetric torsion. We compute their algebra and demonstrate that it closes to a W-algebra upon including additional symmetry generators, one of which is associated with the Nijenhuis form. Due to the presence of chiral fermions in the sigma model action, these symmetries are a priori anomalous. We determine the corresponding chiral anomalies using the Wess--Zumino consistency conditions and establish their consistency at one loop in sigma model perturbation theory. We further show that these anomalies can be cancelled at the same loop order, assuming appropriate quantum corrections to the transformations. Such corrections are compatible with both the cancellation mechanism for spacetime frame rotation and gauge transformation anomalies, and the corrections to heterotic supergravity up to two loops in sigma model perturbation theory. 

\end{abstract}

\vspace{1.5cm}

\newpage

\renewcommand{\thefootnote}{\arabic{footnote}}

\newpage

{
  \hypersetup{linkcolor=blue}
  \tableofcontents
}

\section{Introduction}

It is known \cite{Howe:1991im,Howe:1991vs,Howe:1991ic} following earlier work in this direction \cite{Odake:1988bh,Delius:1989nc} that two-dimensional supersymmetric sigma models with couplings given by a Wess--Zumino term $b$ and a metric $g$ are invariant under a class of transformations generated by covariantly constant forms with respect to a metric connection $\hat{\nabla}$ with skew-symmetric torsion given by the 3-form field strength $H=db$. The target spaces $M$ considered in these investigations are $d$-dimensional manifolds admitting a connection $\hat{\nabla}$ whose holonomy is contained in $U(\frac{d}{2})$, $SU(\frac{d}{2})$, $Sp(\frac{d}{4})$, $Sp(\frac{d}{4})\cdot Sp(1)$, or one of the exceptional groups $G_2$ and $Spin(7)$ in seven and eight dimensions, respectively.  As the presence of these covariantly constant forms implies the reduction of the holonomy group to a proper subgroup of $SO$, the associated symmetries are typically known as holonomy symmetries. Their transformations satisfy a W-algebra \cite{Howe:1991im,Howe:1991vs,Howe:1991ic}, as the structure constants depend on the conserved currents of the theory, and various aspects of these algebras have been investigated at both the classical and quantum level \cite{Howe:1991im, Howe:1991vs,Howe:1991ic,Shatashvili:1994zw,Figueroa-OFarrill:1996tnk,Howe:2010az,Fiset:2021azq,delaOssa:2024cgo}. Some of these backgrounds have also been considered as target spaces of heterotic sigma models, allowing for the investigation of a further aspect of holonomy symmetries, namely their anomalies \cite{deLaOssa:2018iij}.

Indeed, the presence of chiral worldsheet fermions in the heterotic sigma model action renders its symmetries a priori anomalous \cite{Moore:1984ws, Alvarez-Gaume:1985bbj, Bagger:1985pw, Sen:1986mg, Hull:1986xn, Howe:1987nw}. Some of these symmetries are closely related to the underlying geometry of the target space, and the cancellation of their anomalies is therefore necessary to preserve the geometric interpretation of the theory upon quantisation. In particular, the anomalies associated with the symmetries of the gauge sector, namely spacetime frame rotations and gauge transformations, have been investigated in \cite{Hull:1985jv}, where it was shown that their cancellation can be achieved by assigning an appropriate anomalous variation to the Wess--Zumino term $b$ under these transformations. This prescription is compatible with the Green--Schwarz mechanism \cite{Green:1984sg} for the cancellation of the gravitational, gauge and mixed anomalies that arise in supergravity theories. 

Supersymmetric heterotic backgrounds provide a natural setting for the study of holonomy symmetries, as the metric connection $\hat{\nabla}$ with skew-symmetric torsion $H$ arises in the gravitino Killing spinor equation of heterotic supergravity, which is one of the conditions that must be satisfied for the preservation of a fraction of supersymmetry. The existence of Killing spinors further implies that there is a collection of $\hat{\nabla}$-covariantly constant forms of various degrees, constructed as Killing spinor form bilinears \cite{Gran:2005wf}. Thus, when these spacetimes are considered as target spaces of heterotic sigma models, the Killing spinor bilinears can be directly identified as the generators of holonomy symmetries of the theory. The classification of heterotic backgrounds has revealed two distinct classes, depending on whether the holonomy group of $\hat{\nabla}$ is compact or non-compact; see \cite{Gran:2005wf, Gran:2007fu} and the review \cite{Gran:2018ijr}.

The purpose of this work is twofold: first, to determine the algebra of holonomy symmetries, and second, to investigate their chiral anomalies using the Wess--Zumino consistency conditions, and their cancellation at one loop in sigma model perturbation theory. For the latter, we assume that the Killing spinor bilinears remain covariantly constant with respect to the corrected connection. Similar results for heterotic backgrounds with compact holonomy $SU(2)$ and $SU(3)$ can be found in \cite{Grimanellis:2023nav}, while all heterotic backgrounds with non-compact holonomy of the form $K\ltimes\mathbb{R}^8$ have been explored in \cite{Papadopoulos:2023exe}.

To complete the systematic investigation of the holonomy symmetries of heterotic sigma models on supersymmetric heterotic backgrounds with compact holonomy, we focus on the remaining class of backgrounds for which the holonomy of $\hat\nabla$ is contained in $G_2$. Unlike seven-dimensional manifolds with $G_2$ structure, these are ten-dimensional spacetimes that can be locally formulated as principal $G$-bundles over a seven-dimensional conformally balanced Riemannian manifold, with $G$ a three-dimensional Lorentzian Lie group \cite{Gran:2005wf, Gran:2007fu, Gran:2018ijr}. The presence of the additional fibre directions plays a non-trivial role in the commutators of the holonomy symmetry transformations, giving rise to contributions that are absent in sigma models with seven-dimensional target spaces with $G_2$ holonomy. 

The Killing spinor bilinears that generate the holonomy symmetries of the associated sigma models consist of three 1-forms, one 3-form, and one 4-form. The 1-forms are dual to Killing vector fields that generate isometries of spacetime $M$. We show that the closure of the algebra requires three additional symmetry generators. The first involves $TJ_L$, where $T$ is the right-handed energy-momentum tensor generating right-handed worldsheet translations and supersymmetry transformations, and $J_L$ is the conserved current corresponding to the $G_2$-invariant forms $L=\varphi$ and $\psi$. The second symmetry is associated with the quadratic Casimir operator of the Lie algebra of $G$, while the third is generated by the single generically non-vanishing Nijenhuis form of the theory. Remarkably, the latter introduces a new structure in holonomy symmetry algebras that has not been encountered in those previously obtained for the same sigma model on heterotic backgrounds whose holonomy is contained in $SU$.

To explore the chiral anomalies, we use the spacetime frame rotation and gauge transformation anomalies that have been computed in \cite{Zumino1984,Hull:1986xn}, and the Wess--Zumino consistency conditions. The resulting holonomy symmetry anomalies can be expressed in terms of the Chern--Simons form associated with the connection $\check{\nabla}$ with torsion $-H$, ensuring their consistency at one loop. The anomaly cancellation involves the assumption that the corrected Killing spinor bilinears that generate the holonomy symmetries, $L^\hbar$, remain $\hat{\nabla}^\hbar$-covariantly constant with respect to the corrected connection with skew-symmetric torsion $H^\hbar$. This scenario is consistent both with the cancellation of anomalies of the gauge sector symmetries \cite{Hull:1985jv} and the observation that the Killing spinor equations of heterotic supergravity retain their structure up to and including two loops in the sigma model perturbation theory \cite{Bergshoeff:1989de}.

This paper is organised as follows. In Section~\ref{section_2}, we introduce the heterotic sigma model and review its holonomy symmetries, including the general expression for their commutators. We present the formulae for the anomalies and their consistency conditions, and explain how these anomalies can be cancelled. In Section~\ref{section_3}, we describe the geometry of heterotic backgrounds with $G_2$ holonomy and identify the $\hat{\nabla}$-covariantly constant forms that generate the holonomy symmetries of the theory. Furthermore, we express the commutators of these transformations as sums of individual symmetry transformations, determine the holonomy symmetry anomalies, and verify the corresponding Wess--Zumino consistency conditions. In Section~\ref{section_4}, we present our concluding remarks. Finally, in Appendix~\ref{appendix_A}, we provide the identities involving the $G_2$-invariant forms that are useful for the computation of the commutators.

\section{Symmetries of heterotic sigma models}\label{section_2}

\subsection{Action and sigma model symmetries}

The classical fields of a heterotic sigma model with (1,0) worldsheet supersymmetry \cite{Sakamoto:1984zk, Siegel:1983es} are maps $X^\mu$, $\mu=1,\dots,d$, from the superspace $\Xi^{2|1}$ with coordinates $(\sigma^=, \sigma^{\pp} \vert\theta^+)$ into a spacetime $M$, $X:\Xi^{2|1}\rightarrow M$, and Grassmann-odd sections $\Psi_-^\mathrm{a}$, $\mathrm{a}=1,\dots,n$, of a vector bundle $S_-\otimes X^* E$ over $\Xi^{2|1}$, where $S_-$ is the right-handed spinor bundle over $\Xi^{2|1}$ and $E$ a rank-$n$ vector bundle over $M$. The action of the (1,0)-supersymmetric sigma model with couplings $g$, $b$ and $\Omega$ written in terms of (1,0) superfields \cite{Hull:1985jv, Hull:1993ct} is

\begin{equation}\label{het_action}
    S = -i\int \dd \sigma^2 \dd\theta^+ \Big( (g+b)_{\mu\nu}D_+X^\mu\partial_=X^\nu + ih_{\mathrm{a}\mathrm{b}}\Psi_-^\mathrm{a}{\cal D}_+\Psi_-^\mathrm{b} \Big)~.
\end{equation}
Here, $g$ is the spacetime metric and $b$ is a locally defined 2-form on $M$, with $H=db$ a globally defined 3-form. $D_+$ is the superspace derivative with $D^2_+=i\partial_{\pp}$, and the gauge covariant derivative is given by
\begin{equation}
    {\cal D}_+\Psi_-^\mathrm{a} = D_+\Psi_-^\mathrm{a} + D_+X^\mu\Omega_\mu{}^\mathrm{a}{}_\mathrm{b}\Psi_-^\mathrm{b}~,
\end{equation}
where $\Omega$ is the gauge connection on the vector bundle $E$ with curvature $F=d\Omega+ \Omega\wedge \Omega$. The gauge group of the heterotic sigma model is either $Spin(32)/\mathbb
Z_2$ or $E_8\times E_8$, but it can remain arbitrary for our investigation. The fibre metric $h$ on $E$ is used to raise and lower the indices $\mathrm{a,b}$, and without loss of generality\footnote{If not, then $\Omega'_\mu{}^\ma{}_\mb=\Omega_\mu{}^\ma{}_\mb + \frac{1}{2}h^{\ma \mc}{\cal D}_\mu h_{\mc \mb}$ is a metric connection, corresponding to the same action. Locally, one may choose an orthonormal frame, in which the fibre metric takes the form $h_{\ma \mb}=\delta_{\ma\mb}$. Metric compatibility then implies that the connection is skew-symmetric in the fibre indices, $\Omega_{\mu \ma\mb}=-\Omega_{\mu \mb\ma}$.
}, we can consider $\Omega$ to be a metric connection, ${\cal D}_\mu h_{\ma \mb}=0$. This action is manifestly invariant under (1,0) worldsheet supersymmetry transformations. 

The superfields can be written in terms of components as
\begin{equation}
    X^\mu(\sigma,\theta^+)=x^\mu(\sigma)+\theta^+\lambda^\mu_+(\sigma)~, \quad \Psi_-^\mathrm{a}(\sigma,\theta^+) = \psi_-^\mathrm{a}(\sigma) + \theta^+F^\mathrm{a}(\sigma)~, 
\end{equation}
where $x^\mu$ are scalar fields and $\theta^+$ is a right-handed Majorana--Weyl worldsheet spinor. It then follows that $\lambda^\mu_+$ are left-handed Majorana--Weyl spinors, while $\Psi^\mathrm{a}_-$ and $\psi^\mathrm{a}_-$ are right-handed, and $F^\mathrm{a}$ are auxiliary scalar fields. 

Varying the action (\ref{het_action}) one finds
\begin{equation}\label{S_variation}
    \delta S = -i\int \dd^2\sigma \dd\theta^+ \left( \delta X^\mu {\cal S}_\mu + \Delta\Psi^\ma_-{\cal S}_\ma \right)~,
\end{equation}
where 
\begin{equation}
    \Delta\Psi^\ma_- \coloneqq \delta\Psi^\ma_- + \delta X^\mu\Omega_\mu{}^\ma{}_\mb \Psi^\mb_-~,
\end{equation}
is the covariant variation of $\Psi_-$ and
\begin{align}\label{field_eqn}
    {\cal S}_\mu &=-2g_{\mu\nu}\hat{\nabla}_= D_+X^\nu - i\Psi^\ma_-\Psi^\mb_-D_+X^\nu F_{\mu\nu\ma\mb}~, \nn
    {\cal S}^\ma &=2i{\cal D}_+\Psi_-^\ma~, 
\end{align}
determine the field equations ${\cal S}_\mu=0$ and ${\cal S}_\ma=0$. The connection $\hat{\nabla}$ is given by
\begin{equation}
    \hat{\nabla}_\nu Y^\mu = \nabla_\nu Y^\mu + \frac{1}{2}H^\mu{}_{\nu\rho}Y^\rho~,
\end{equation}
where $\nabla$ is the Levi--Civita connection on $M$ with respect to the spacetime metric $g$, and $H$ is the skew-symmetric torsion of $\hat{\nabla}$.

The sigma model symmetries are those transformations that leave the action invariant by changing both the fields $X$ and $\Psi_-$ and the couplings $g$, $b$, $h$, and $\Omega$. These are formal symmetries in the sense that they do not correspond to a Noether current, unlike the standard symmetries of the action, which involve variations of the fields $X$ and $\Psi_-$ alone. These are the diffeomorphisms, the gauge transformations, and the spacetime frame rotations. Infinitesimal diffeomorphisms generated by the vector field $v$ yield the following transformations for the fields and the coupling constants
\begin{align}\label{diffeo_transf}
    &\delta X^\mu=v^\mu~,\quad \delta g_{\mu\nu} = -{\cal L}_v g_{\mu\nu}~, \quad \delta b_{\mu\nu} = -{\cal L}_v b_{\mu\nu}~, \quad \delta\Psi_-^\ma = w^\ma{}_\mb\Psi_-^\mb~, \nn
    &\delta\Omega_\mu{}^\ma{}_\mb = -{\cal L}_v\Omega_\mu{}^\ma{}_\mb- D_\mu^\Omega w^\ma{}_\mb ~, \quad \delta h_{\ma\mb} = - {\cal L}_v h_{\ma\mb} - w^\mc{}_\ma h_{\mc\mb} - h_{\ma\mc}w^\mc{}_\mb~,
\end{align}
where $D_\mu^{\cal A}Z^\ma{}_\mb=\partial_\mu Z^\ma{}_\mb + {\cal A}_\mu{}^\ma{}_\mc Z^\mc{}_\mb - {\cal A}_\mu{}^\mc{}_\mb Z^\ma{}_\mc$ for a connection ${\cal A}$, and it is assumed that the diffeomorphism lifts to the bundle $E$, inducing a fibre rotation parametrised by $w(v)$. These fibre rotations act as gauge transformations, which we consider independently below, and thus we can set $w(v)=0$. The infinitesimal gauge transformations, which are symmetries of the gauge sector of (\ref{het_action}), are parametrised by $u$ with the following variations
\begin{align}\label{gauge_transf}   
    \delta_u\Psi_-^\ma = u^\ma{}_\mb\Psi_-^\mb~, \quad \delta_u\Omega_\mu{}^\ma{}_\mb = -D^\Omega_\mu u^\ma{}_\mb~, \quad \delta_u h_{\ma\mb} = -u^\mc{}_\ma h_{\mc\mb} - h_{\ma\mc}u^\mc{}_\mb~.
\end{align}
    
For the purpose of quantisation, it is convenient to introduce a pseudo-orthonormal frame $e_\mu^A$ on the tangent bundle of spacetime, allowing the quantum fluctuations to be expressed in the frame basis when applying the background field method discussed in Section \ref{section_WZ_ACC}. Writing the spacetime metric as $g_{\mu\nu}=\eta_{AB}e_\mu^A e_\nu^B$, the action (\ref{het_action}) remains invariant under infinitesimal spacetime frame rotations, parametrised by $\ell$

\begin{equation}\label{frame_transf}
\delta_\ell e_\mu^A=\ell^A{}_B e_\mu^B~,
\quad
\delta_\ell\omega_\mu{}^A{}_B=-D^\omega_\mu\ell^A{}_B~,
\end{equation}
where $\omega_\mu{}^A{}_B$ is the spin connection of the tangent bundle that preserves the spacetime metric, $\omega_{\mu AB}=-\omega_{\mu BA}$. Under diffeomorphisms, $\omega$ transforms in a manner similar to $\Omega$. The commutator of two spacetime frame rotations is $[\delta_\ell,\delta_{\ell'}]=\delta_{[\ell,\ell']}$, where $[\cdot,\cdot]$ is the matrix commutator, and similarly for the commutator of the gauge transformations (\ref{gauge_transf}).

The coupling $b$ is defined locally in patches, differing from patch to patch by a gauge transformation 
\begin{equation}
    \delta b_{\mu\nu} = (dm)_{\mu\nu}~, 
\end{equation}
for a 1-form $m$ on the spacetime, which also leaves the action (\ref{het_action}) invariant.

\subsection{Holonomy symmetries}

Another class of symmetries is given by the transformations

\begin{equation}\label{hol_transf}
    \delta_L X^\mu =  a_L L^\mu{}_{\cal L} D_+X^{\cal L}~, \quad \Delta_L \Psi^\ma_- = 0~,
\end{equation}
where the multi-index $\cal L$ denotes a skew-symmetrised string of $\ell$ indices, ${\cal L}\coloneqq \lambda_1\dots\lambda_\ell$, and $D_+X^{\lambda_1\dots \lambda_\ell}$ is a shorthand notation for $D_+X^{\lambda_1}\dots D_+X^{\lambda_\ell}$. The parameters $a_L$ have Grassmann parity $(-1)^\ell$ and Lorentz weight $-\ell/2$, chosen such that the transformations $\delta_L X$ are Grassmann even. This transformation is non-linear in the superspace derivatives of $X^\mu$ for $\ell \geq 2$, and it is an off-shell symmetry of the theory \cite{Howe:1991im,Howe:1991vs,Howe:1991ic}, provided that it is semi-local with $\partial_=a_L=0$, i.e., $a_L$ does not depend on all superspace coordinates, $a_L = a_L(\sigma^{\pp},\theta^+)$, and $L$ satisfies 
\begin{equation}\label{inv_cond}
    \hat{\nabla}_\mu L_{\lambda_1\dots \lambda_{\ell+1}} = 0 ~, \quad F_{\nu[\lambda_1}L^\nu{}_{\lambda_2\dots\lambda_{\ell+1}]} = 0~.
\end{equation}

Such forms arise naturally in heterotic backgrounds as bilinears of Killing spinors that satisfy the gravitino and gaugino Killing spinor equations 
\begin{equation}\label{het_KSE_1}
    \hat{\nabla}_\mu\epsilon=0~, \quad F_{\mu\nu}\Gamma^{\mu\nu}\epsilon = 0~,
\end{equation}
which together with the dilatino Killing spinor equation
\begin{equation}\label{het_KSE_2}
    \left(\Gamma^\mu\partial_\mu\Phi -\frac{1}{12}H_{\mu\nu\rho}\Gamma^{\mu\nu\rho} \right)\epsilon = 0~,
\end{equation}
determine the conditions that heterotic backgrounds must satisfy to preserve a fraction of the original spacetime supersymmetry $N=16$. In general, given a set of $\hat{\nabla}$-covariantly constant spinors $\epsilon_r$, the bilinears associated with the bispinors $\epsilon_r\otimes\epsilon_s$ are also $\hat{\nabla}$-covariantly constant. Such bilinears can be written as $Spin$-invariant inner products involving skew-symmetrised products of gamma matrices, schematically taking the form $\tau_{rs}=\langle \epsilon_r, \Gamma\dots\Gamma \epsilon_s \rangle$. See also Appendix B in \cite{Gran:2018ijr} for their general properties and \cite{Gran:2005wf} for those appearing in heterotic backgrounds classified by the holonomy of $\hat{\nabla}$. 

A straightforward example of a symmetry of the form (\ref{hol_transf}) is that generated by a Killing vector field $L=K$ for $\ell=0$ with $i_KH=dK^\flat$. A generalisation of the transformation (\ref{hol_transf}) which includes the gauge sector and a potential term\footnote{This is of the form $\mu s_\ma \Psi^\ma_-$ where the parameter $\mu$ has mass dimensions and $s_\ma$ is a section of $E^*$ \cite{Hull:1993ct}.} has been given in \cite{deLaOssa:2018iij}, which assigns a non-vanishing covariant variation to $\Psi_-$. 

The existence of $\hat{\nabla}$-covariantly constant forms restricts the holonomy of $\hat{\nabla}$ to a subgroup of $SO$ that leaves these forms invariant. It is for this reason that these transformations (\ref{hol_transf}) are typically referred to as holonomy symmetries\footnote{They should not be confused with transformations of $\mathrm{hol}(\hat\nabla)$ group which are not symmetries of the theory.}. The restriction of the holonomy can also be understood through the integrability condition 
\begin{equation}
    [\hat{\nabla}_\mu,\hat{\nabla}_\nu]L_{\lambda_1\dots \lambda_{\ell+1}} =0~,
\end{equation}
which implies
\begin{equation}\label{RL_eqn}
    \hat{R}_{\rho[\lambda_1,|\mu\nu|}L^\rho{}_{\lambda_2\dots \lambda_{\ell+1}]}=0~.
\end{equation} 
Here $\hat R^\rho{}_{\lambda,\mu\nu}$ is the curvature tensor of $\hat\nabla$; see Appendix \ref{appendix_A} for our conventions. Since the curvature takes values in the holonomy algebra, $\hat{R}_{\mu\nu}\in\mathfrak{hol}(\hat\nabla)$, equation (\ref{RL_eqn}) expresses the invariance of $L$ under the holonomy group action.

The possible holonomy groups of $\hat{\nabla}$ for heterotic backgrounds in which all $\hat{\nabla}$-parallel spinors are Killing have been classified in \cite{Gran:2005wf, Gran:2007fu}, and the bilinears of each case have been found; see also the review \cite{Gran:2018ijr}. These groups fall into two classes: compact 
\begin{equation}
    G_2~ (2)~, \quad SU(3)~ (4)~, \quad SU(2)~ (8)~, \quad \lbrace 1 \rbrace~(16)~,
\end{equation}
and non-compact $K\ltimes  \mathbb{R}^8$, with $K$ being one of the following groups 
\begin{align}\label{non-compact_hol_groups}
    &Spin(7) ~ (1)~, \quad SU(4) ~ (2)~, \quad Sp(2)~ (3)~, \quad \times^2 Sp(1)~ (4)~, \quad \nn &Sp(1) ~ (5)~, \quad U(1) ~ (6)~, \quad \lbrace 1 \rbrace~ (8)~. 
\end{align}
The numbers in parentheses denote the number of Killing spinors for each background. 

The commutator of two  transformations (\ref{hol_transf}) on the field $X$ generated by the vector-valued $\ell$ and $m$-forms $L$ and $M$, respectively, can be written \cite{Howe:1991ic} as
\begin{align}\label{general_hol_commut}
[\delta_L, \delta_M]X^\mu= \delta_{LM}^{(1)} X^\mu+\delta_{LM}^{(2)} X^\mu+\delta_{LM}^{(3)} X^\mu~,
\end{align}
with
\begin{align}\label{LM_term_1}
\delta_{LM}^{(1)} X^\mu=a_M a_L N(L,M)^\mu{}_{{\cal L}{\cal M}} D_+X^{{\cal L}{\cal M}}~,
\end{align}
\begin{align}\label{LM_term_2}
\delta_{LM}^{(2)} X_\mu=\big (&-m a_M D_+a_L (L\cdot M)_{\nu {\cal L}_2, \mu {\cal M}_2} \nn
&+ \ell (-1)^{(\ell+1) (m+1)} a_L D_+ a_M (L\cdot M)_{\mu {\cal L}_2,\nu {\cal M}_2}\big)
 D_+X^{\nu {\cal L}_2{\cal M}_2}~,
\end{align}
\begin{align}\label{LM_term_3}
\delta_{LM}^{(3)} X_\mu=-2i \ell m (-1)^\ell a_M a_L (L\cdot M)_{(\mu|{\cal L}_2|, \nu ){\cal M}_2} \partial_{\pp}X^\nu D_+X^{{\cal L}_2{\cal M}_2}~,
\end{align}
where $(L\cdot M)_{\lambda {\cal L}_2,\mu {\cal M}_2}=L_{\rho \lambda[{\cal L}_2} M^\rho{}_{|\mu| {\cal M}_2]}$, and ${\cal L}_2$ and ${\cal M}_2$ denote the skew-symmetrised string of indices ${\cal L}_2\coloneqq\lambda_2\dots\lambda_\ell$ and ${\cal M}_2\coloneqq\mu_2\dots \mu_m$. Moreover, the associated current
\begin{equation}
    J_L = L_{\lambda_1\dots\lambda_{\ell+1}}D_+X^{\lambda_1\dots \lambda_{\ell+1}}~,
\end{equation}
is conserved, $\partial_=J_L=0$, subject to the field equations (\ref{field_eqn}). The Nijenhuis tensor \cite{Frolicher:1956nk1,Nijenhuis:1955dl, Nijenhuis:1955dm} of the vector forms $L$ and $M$ that appears in (\ref{LM_term_1}) is given by the expression
\begin{align}
N(L,M)^\mu=\Big (L^\nu{}_{\cal L}\partial_\nu M^\mu{}_{\cal M}- M^\nu{}_{\cal M} \partial_\nu L^\mu{}_{\cal L} -\ell L^\mu{}_{\nu {\cal L}_2} \partial_{\lambda_1} M^\nu{}_{\cal M} +m M^\mu{}_{\nu {\cal M}_2} \partial_{\mu_1} L^\nu{}_{\cal L}\Big) dx^{{\cal L}{\cal M}}~,
\end{align}
and it is a type--$(1,\ell+m)$ tensor that reduces to the standard Nijenhuis tensor for $L=M=I$, where $I$ is an almost complex structure. The vanishing of $N(I,I)$ is equivalent to the integrability of $I$, in which case $I$ defines a complex structure. In heterotic backgrounds, such cases appear when $\text{hol}(\hat{\nabla})$ is contained in $SU(3)$ or $SU(2)$. The former case is associated with K\"ahler with torsion geometry and the latter with hyper-K\"ahler with torsion geometry, giving rise to $(2,0)$ and $(4,0)$ worldsheet supersymmetry, respectively. However, no such complex structures arise in $G_2$ backgrounds, and thus no additional worldsheet supersymmetry appears. 

Using that $L$ and $M$ are $\hat{\nabla}$-covariantly constant, the Nijenhuis tensor of $L$ and $M$ can be further written as 
\begin{align}\label{Nijenhuis_tensor}
    N(L,M)_{\mu{\cal L}{\cal M}} ={}&- H_{\mu\nu\rho}L^\nu{}_{\cal L} M^\rho{}_{\cal M}
    + m L^\nu{}_{\cal L}H_{\nu\mu_1}{}^\rho M_{\mu\rho {\cal M}_2} \nn
    &- \ell M^\nu{}_{\cal M}H_{\nu\lambda_1}{}^\rho L_{\mu\rho {\cal L}_2}
    + \ell mH^\rho{}_{\lambda_1\mu_1}
    (L\!\cdot\! M)_{(\mu|{\cal L}_2|,\rho){\cal M}_2} \nn
    ={}&-(\ell+m+1) H_{[\mu|\nu\rho|}L^\nu{}_{\cal L}M^\rho{}_{{\cal M}]}
    + \ell mH^\rho{}_{\lambda_1\mu_1}
    (L\!\cdot\! M)_{(\mu|{\cal L}_2|,\rho){\cal M}_2}~,
\end{align}
and therefore vanishes for the torsion-free connection. From the last equality, it can be seen that the first term is totally skew-symmetric, but not $\hat{\nabla}$-covariantly constant. 

A totally skew-symmetric and $\hat{\nabla}$-covariantly constant Nijenhuis form has been constructed in \cite{Howe:2010az} and defined in the total space. However, as we shall see, heterotic backgrounds with compact holonomy are locally modelled as principal bundles, with the invariant forms associated with the holonomy group defined on the base space. Thus, for our investigation, we shall adjust the treatment of \cite{Howe:2010az} accordingly. One can then construct from $L$ and $M$ an ($\ell+m$)-form $P$ and an $(\ell+m-2)$-form $Q$ that satisfy \cite{Howe:2006si}
\begin{align}\label{PQ_eqn}
(L \cdot M)_{i[\bar{{\cal L}}_2,j\bar{{\cal M}}_2]}
&=
(-1)^{\ell+1} P_{ij\bar{{\cal L}}_2\bar{{\cal M}}_2}
+\frac{m}{2} g_{i[j} Q_{\bar{{\cal L}}_2\bar{{\cal M}}_2]}~, \nn
(L \cdot M)_{[j\bar{{\cal L}}_2,|i|\bar{{\cal M}}_2]}
&=
(-1)^\ell P_{ij\bar{{\cal L}}_2\bar{{\cal M}}_2}
+\frac{\ell}{2}g_{i[j} Q_{\bar{{\cal L}}_2\bar{{\cal M}}_2]}~, \nn
(L \cdot M)_{i[\bar{{\cal L}}_2,|j|\bar{{\cal M}}_2]}
+ (i \leftrightarrow j)
&=
g_{ij} Q_{\bar{{\cal L}}_2\bar{{\cal M}}_2}
-(\ell+m-2) g_{(i[l_2} Q_{j)\bar{{\cal L}}_3\bar{{\cal M}}_2]}~,
\end{align}
where all indices are taken to be along the base space, including ${\cal \bar{L}}\coloneqq l_1\dots l_\ell$ and ${\cal \bar{M}}\coloneqq m_1\dots m_m$. Then the Nijenhuis form of degree $\ell+m+1$ reads
\begin{align}\label{Nijen_form}
{\cal N}_{i{\cal \bar{L} \bar{M}}}
=
-(\ell+m+1)
\left(
H_{jk[i}L^j{}_{\bar{{\cal L}}} M{}^k{}_{\bar{{\cal M}}}
+
(-1)^\ell \frac{\ell m}{6}
H_{[i l_1 l_2}
Q_{\bar{{\cal L}}_3 \bar{{\cal M}}]}
\right)~,
\end{align}
and the transformation with parameter $a_{\cal N}$ proportional to $a_La_M$ and Lorentz weight $-(\ell+m)/2$ 
\begin{equation}\label{nijen_form_transf}
    \delta_{{\cal N}(L,M)} X^i = a_{{\cal N}} {\cal N}(L,M)^i{}_{\cal \bar{L} \bar{M}} D_+X^{\cal \bar{L} \bar{M}}~, \quad \Delta_{{\cal N}(L,M)}\Psi^\ma_-=0~,
\end{equation}
can be a symmetry of the theory\footnote{The proof of (\ref{nijen_form_transf}) being a symmetry for any $\hat{\nabla}$-covariantly constant $L$ and $M$ in \cite{Howe:2010az} follows from the Bianchi identity (\ref{Bianchi1}) for a closed 3-form field strength. However, since here ${\cal N}$ is horizontal, this would require the horizontal component of $H$ to be closed, which is not generally the case.} under a specific choice of $L$ and $M$, which we shall explore in Section \ref{commutators}. An additional symmetry generated by the vector $(q+1)$-form 
\begin{equation}
    S^\mu{}_{\nu\rho_1\dots\rho_q} = g^{\mu\sigma}S_{\sigma,\nu\rho_1\dots\rho_q} = \delta^\mu{}_{[\nu}Q_{\rho_1\dots\rho_q]}~,
\end{equation}
with the $\hat{\nabla}$-covariantly constant $q$-form $Q$ defined by (\ref{PQ_eqn}) with $q=\ell+m-2$, is given by \cite{Howe:2010az}
\begin{align}\label{S_transf}
    \delta_S X^\mu &= \alpha_S \hat{\nabla}_+D_+X^\nu S_{\nu,\mu {\cal Q}}D_+X^{\cal Q}
    + \frac{(-1)^q}{q+1}\hat{\nabla}_+ \left( \alpha_S S_{\mu,\nu{\cal Q}} D_+X^{\nu{\cal Q}} \right) \nn
    &\hphantom{={}}- \frac{q+3}{3(q+1)}\alpha_S H_{[\mu\nu\rho}Q_{\cal Q]} D_+X^{\nu\rho{\cal Q}}~, \nn
    [0.4cm]
    \Delta_S \Psi^\ma_- &= -\frac{(-1)^q}{q+1} \alpha_S Q_{\cal Q}F_{\mu\nu}{}^{\ma}{}_{\mb}\Psi^{\mb}_- D_+X^{{\cal Q}\mu\nu}~,
\end{align}
which, unlike the holonomy symmetry transformations, has a non-vanishing covariant transformation on $\Psi_-$. When writing the commutator of two holonomy symmetries as a sum of individual symmetries, one may encounter transformations of the form (\ref{S_transf}). It is also straightforward to prove that it leaves the action (\ref{het_action}) invariant by using the Bianchi identity (\ref{Bianchi1}).

The special case $Q=1$ corresponds to the transformations generated by the right-handed energy-momentum tensor of the theory
\begin{equation}
    T_{+\pp} = g_{\mu\nu} D_+X^\mu\hat{\nabla}_+D_+X^\nu - \frac{1}{3}H_{\mu\nu\rho}D_+X^{\mu\nu\rho}~,
\end{equation}
where the subscripts denote the superspace directions. The corresponding transformation is given by
\begin{equation}
    \delta_TX^\mu = 2i\alpha_T\partial_{\pp}X^\mu + D_+\alpha_T D_+X^\mu~, \quad \Delta_T\Psi^\ma_- = -\alpha_T F_{\mu\nu}{}^\ma{}_\mb\Psi^\mb_-D_+X^{\mu\nu}~, 
\end{equation}
where the first term of $\delta_T$ represents right-handed worldsheet translations, and the second term represents right-handed supersymmetry transformations. The current associated with (\ref{S_transf}) is $J_S=T_{+\pp}J_Q$, and it is conserved, $\partial_=J_S=0$, as a result of the conservation of the energy-momentum tensor, $\partial_=T_{+\pp}=0$ subject to the field equations (\ref{field_eqn}), and the Bianchi identity (\ref{Bianchi1}).

Focusing, finally, on the commutator on $\Psi_-$, given that $\Delta_L\Psi_-=\Delta_M\Psi_-=0$, one obtains   
\begin{equation}\label{psi_commutator}
    [\delta_L,\delta_M]\Psi^{\ma}_- = - F_{\mu\nu}{}^\ma{}_\mb \delta_LX^\mu\delta_M X^\nu\Psi^\mb_- - \Omega_\mu{}^\ma{}_\mb[\delta_L,\delta_M]X^\mu\Psi^\mb_-~.
\end{equation}
The first term gives a non-trivial covariant transformation, while the second term depends on the commutator on $X$. Thus, we have
\begin{equation}\label{Psi_term}
    \Delta_{LM}\Psi^\ma_- = - F_{\mu\nu}{}^\ma{}_\mb \delta_L X^\mu \delta_M X^\nu \Psi^\mb_-~.
\end{equation}
Such a term is also necessary for consistency with the non-vanishing $\Delta_S\Psi_-$ transformation (\ref{S_transf}). Thus, the algebra is represented on the full multiplet $(X,\Psi_-)$, rather than on $X$ alone. However, it suffices from the general expression (\ref{psi_commutator}) to focus on the algebra on $X$ and indicate in each case the term (\ref{Psi_term}).

It can always be arranged for the target space diffeomorphisms (\ref{diffeo_transf}) to be non-anomalous by adding a finite local counterterm to the effective action, which shifts the corresponding anomaly to the anomaly of the spacetime frame rotations \cite{Alvarez-Gaume:1985bbj, BARDEEN1984421, Alvarez-Gaume:1984zlq}. Thus, we shall focus on the algebra of the holonomy symmetries, as one can readily verify that the commutators of the gauge sector transformations (\ref{gauge_transf}) and (\ref{frame_transf}) with the holonomy symmetry transformations (\ref{hol_transf}) vanish
\begin{equation}\label{gauge_frame_hol_commut}
    [\delta_\ell,\delta_L]=0~, \quad [\delta_u,\delta_L]=0~.
\end{equation}
To compute the commutator (\ref{general_hol_commut}), we shall rearrange the right-hand side as a sum of individual symmetries of the action. This will also be useful for analysing the anomaly consistency conditions, which require that the anomaly be compatible with the symmetry algebra. Examples of this approach are found in \cite{Howe:2006si} and \cite{Howe:2010az} for vanishing and non-vanishing $H$, respectively. Note, though, that the conditions for the $\hat{\nabla}$-covariantly constant forms in those cases do not coincide with the corresponding conditions in heterotic supersymmetric backgrounds with compact holonomy, since the latter are fibrations with a fibre group manifold; see Section \ref{section_the_G2_geometry} and references therein. The conditions in \cite{Howe:2006si, Howe:2010az} generally do not hold along the fibre directions, but they are still valid if we restrict the right-hand side of the commutator (\ref{general_hol_commut}) to the base space directions. We shall therefore incorporate this treatment into our more general setting and show that the commutator can be expressed as a sum of individual symmetries along all spacetime directions, including additional symmetries identified in each case separately.

\subsection{Quantisation and anomaly consistency conditions}\label{section_WZ_ACC}

A convenient approach to quantising the theory and investigating its anomalies is the background field method \cite{Honerkamp:1971sh, Alvarez-Gaume:1981exa, Friedan:1980jm}, where the total field is split into a classical background field and a quantum fluctuation. To preserve covariance, the splitting is defined through geodesics, so that the quantum fluctuations transform as tangent vectors. It is then convenient to express these fluctuations in an pseudo-orthonormal frame, in which their kinetic terms take a standard form. When the effective action $\Gamma$ is expressed in terms of one-particle irreducible diagrams with external background fields, one can distinguish two classes of symmetries. Spacetime frame rotations and gauge transformations act linearly on the quantum fields \cite{Howe:1987nw}, unlike holonomy symmetries that need further investigation \cite{Howe:2006si}.

In this Section, we first state the frame rotation and gauge anomalies in terms of the background fields and use them in the consistency conditions to derive the anomalies of the holonomy symmetries. From this point onward, all fields are treated as background fields. 

As the symmetry transformations of the classical theory $\delta_A$ and $\delta_B$ generated by $A$ and $B$, respectively, close under the commutation relation 
\begin{equation}
    [\delta_A,\delta_B] = \delta_{[A,B]}~,
\end{equation}
any quantum violation of these symmetries, i.e. $\delta_A\Gamma=\Delta(a_A)$ with $a_A$ the transformation parameter, must remain compatible with the underlying symmetry algebra. Applying the above relation to the effective action $\Gamma$ immediately gives the Wess--Zumino anomaly consistency conditions \cite{Wess:1971yu}
\begin{equation}\label{WZ_CC}
    \delta_A\Delta(a_B) - \delta_B\Delta(a_A) = \Delta(a_{[A,B]})~.
\end{equation}
Anomalies that violate these conditions would immediately signal an inconsistency of quantum theory. 

To write down the anomaly associated with the spacetime frame rotations, one can use the descent equations \cite{Zumino1984} and start with the first Pontryagin form of the manifold, $P_4= \text{tr}\left( R(\omega)\wedge R(\omega)\right)$, where $R(\omega)=d\omega+\omega\wedge\omega$ is the curvature of a connection $\omega$. Using the Bianchi identity $DR=dR+[\omega,R]=0$, one can see that $P_4$ is closed. The Poincar\'e lemma then implies that $P_4$ is locally exact, $P_4=dQ^0_3(\omega)$, where $Q^0_3(\omega)$ is the Chern--Simons 3-form. Since, by construction, $P_4$ is invariant under spacetime frame rotation transformations, one has $d\delta_\ell Q^0_3(\omega)=0$, and the variation of the Chern--Simons form is also locally exact
\begin{equation}
    \delta_\ell Q^0_3(\omega) = dQ^1_2(\omega,\ell)~,
\end{equation}
where $Q^1_2$ is linear in $\ell$. The anomaly is then identified with the integrated descendant
\begin{equation}\label{frame_anomaly}
    \Delta(\ell) = \frac{i\hbar}{4\pi} \int \dd^2\sigma\dd\theta^+ Q^1_2(\omega,\ell)_{\mu\nu} D_+X^\mu\partial X_=^\nu ~,
\end{equation}
where the numerical coefficient can be determined from topological arguments \cite{Zumino1984,Atiyah:1984tf, Alvarez-Gaume:1983ict}, Fujikawa's method \cite{Alvarez-Gaume:1983ict} or explicit diagrammatic calculations \cite{Adler:1969gk, Bell:1969ts, Bardeen:1969md}. Similarly, the gauge anomaly can be written as
\begin{equation}
    \Delta(u) = -\frac{i\hbar}{4\pi} \int \dd^2\sigma\dd\theta^+ Q^1_2(\Omega,u)_{\mu\nu} D_+X^\mu\partial X_=^\nu ~,
\end{equation}
where the relative minus sign is due to the opposite chiralities of the fermions. The left-handed fermions $\lambda^\mu_+$ couple to the target space frame connection, while the right-handed fermions $\psi^\ma_-$ couple to the gauge connection. Note that the connections appearing in the above expressions are not uniquely fixed \cite{Zumino1984} and can be chosen such that the Wess--Zumino consistency conditions are satisfied. The spacetime frame rotations and gauge transformations act on different bundles, $[\delta_u,\delta_\ell]=0$, and therefore the consistency condition becomes $\delta_u\Delta(\ell)-\delta_\ell\Delta(u)=0$, which is trivially satisfied since $\Omega$ and $\omega$ are invariant under $\delta_\ell$ and $\delta_u$ transformations, respectively. 

As the commutator of holonomy symmetries with spacetime frame rotations and gauge transformations vanishes, $[\delta_\ell,\delta_L]=[\delta_u,\delta_L]=0$, we can use (\ref{WZ_CC}) for both cases
\begin{equation}
    \delta_\ell\Delta(a_L) - \delta_L\Delta(\ell) = 0~, \quad 
    \delta_u\Delta(a_L) - \delta_L\Delta(u) = 0~,
\end{equation}
to calculate the corresponding holonomy anomaly
\begin{equation}\label{hol_anomaly}
    \Delta(a_L) = \frac{i\hbar}{4\pi}\int \dd^2\sigma \dd\theta^+ Q^0_3(\omega,\Omega)_{\mu\nu\rho} \delta_LX^\mu D_+X^\nu \partial_= X^\rho + \Delta_{\mathrm{inv}}(a_L)~,
\end{equation}
up to terms invariant under both $\delta_\ell$ and $\delta_u$, collectively denoted by $\Delta_{\mathrm{inv}}(a_L)$, where $Q^0_3(\omega,\Omega)=Q^0_3(\omega)-Q^0_3(\Omega)$. At the loop order considered here, we assume that $\Delta_{\rm inv}(a_L)$ vanishes. The possibility of additional invariant contributions at higher orders in sigma model perturbation theory is discussed in \cite{deLaOssa:2018iij,Grimanellis:2023nav}.

For the special case of a Killing vector $L=K$ satisfying $i_K P_4=0$, the above expression simplifies. In particular, ${\cal L}_K P_4=0$ and locally ${\cal L}_K Q^0_3(\omega)=dQ^1_2(\omega,a_K)$. Using the descent equations and Cartan's identity for the Lie derivative, one obtains
\begin{equation}
    {\cal L}_KQ^0_3(\omega) = di_KQ^0_3(\omega) + i_KP_4~,
\end{equation}
and hence $i_KQ^0_3(\omega) = Q^1_2(\omega,a_K)$ up to a closed 2-form, which can be absorbed into a redefinition of $Q^1_2(\omega,a_K)$. Consequently, the anomaly can be written as
\begin{equation}
    \Delta(a_K) = \frac{i\hbar}{4\pi}\int \dd^2\sigma\dd\theta^+ Q^1_2(\omega,a_K)_{\mu\nu} D_+X^\mu \partial_=X^\nu~,
\end{equation}
which, using (\ref{KV_algebra}), can readily be shown to satisfy the consistency conditions (\ref{WZ_CC}). The remaining consistency conditions involve the commutator of two holonomy symmetries generated by $L$ and $M$ 
\begin{align}\label{WZ_CC_hol}
    \delta_L\Delta(a_M) - \delta_M\Delta(a_L) &= \frac{i\hbar}{4\pi}\int \dd^2\sigma\dd\theta^+ P_4(R,F)_{\mu\nu\rho\sigma}\delta_LX^\mu \delta_MX^\nu D_+X^\rho \partial_=X^\sigma \nn
    & + \frac{i\hbar}{4\pi}\int \dd^2\sigma\dd\theta^+ Q^0_3(\omega,\Omega)_{\mu\nu\rho}[\delta_L,\delta_M]X^\mu D_+X^\nu \partial_=X^\rho~.
\end{align}
Comparing (\ref{WZ_CC_hol}) with (\ref{hol_anomaly}), one may notice a potential inconsistency, as there is an additional term on the right-hand side of the former, proportional to $P_4(R,F)=P_4(R)-P_4(F)$. The consistency condition will also depend on the individual symmetries that appear in the decomposition of the commutator $[\delta_L,\delta_M]$. If none of those symmetries is anomalous, then the whole right-hand side of (\ref{WZ_CC_hol}) must vanish for consistency.

\subsection{Anomaly cancellation and consistency conditions revisited}\label{section_anomalies}

Quantum corrections are expected to preserve both the holonomy of the torsionful connection and the number of Killing spinors. However, the classical description of the symmetries may have to be modified, with two natural possibilities at one loop. 

The first method is that an anomaly can be removed by adding a finite local counterterm to the effective action in a renormalisation scheme that leaves the symmetry generators uncorrected. As $P_4$ is locally exact, there is freedom to choose $Q^0_3$ up to an exact form, $Q^0_3\rightarrow Q^0_3+dW$, which can be exploited to construct such counterterms from a 2-form $W$. A successful example is the cancellation of anomalies associated with additional worldsheet supersymmetries generated by complex structures \cite{Sen:1986mg, Howe:1987nw}. In the $SU(2)$ and $SU(3)$ holonomy cases \cite{Grimanellis:2023nav} the hyper-K\"ahler geometry and the local triviality of the Bott--Chern cohomology of the corresponding base spaces, respectively, allow for the decomposition of the Chern--Simons form into appropriate terms. These involve the complex structures and exact forms, and they are used in the construction of finite local counterterms for the anomaly cancellation. Nevertheless, we cannot adopt this approach in our investigation, as the seven-dimensional base space of heterotic backgrounds with $G_2$ holonomy does not admit a complex structure. However, finite local counterterms can still be used to cancel the anomalies associated with the Killing vectors $K$, following the construction in \cite{Grimanellis:2023nav}, and thus $K$ does not receive quantum corrections at this order.

The second method, which is the most suitable for our work, is to consider quantum corrections to the bilinears $L$, denoted by $L^\hbar$. Combining the variation (\ref{S_variation}) under the corrected transformation with the anomaly (\ref{hol_anomaly}), one obtains
\begin{align}\label{corrected_anomaly}
    \delta_L^\hbar\Gamma = -i\int \dd^2\sigma\dd\theta^+ \Big(&a_L \frac{2(-1)^\ell}{\ell+1}\hat{\nabla}^\hbar_\mu L^\hbar_{{\cal L}+1}\partial_= X^\mu D_+X^{{\cal L}+1}\nn 
    &- ia_L L^{\hbar\mu}{}_{\cal L}F^\hbar_{\mu\nu\ma\mb}\Psi^\ma_-\Psi^\mb_- D_+X^{{\cal L}\nu} + 2i\Delta_L^\hbar\Psi^\ma_-{\cal D}^\hbar_+\Psi_{-\ma} \Big) + {\cal O}(\hbar^2)~,
\end{align}
where $\hat{\nabla}^\hbar$ is the quantum-corrected connection with torsion\footnote{One should also expect $g$ and $b$ to receive quantum corrections, though for simplicity we shall drop the $\hbar$ superscript.} 
\begin{equation}\label{H_corrected}
    H^\hbar = H -\frac{\hbar}{4\pi}Q^0_3(\omega,\Omega) + {\cal O}(\hbar^2)~.
\end{equation}
Moreover, ${\cal D}^\hbar$ is the corrected connection of the gauge sector, and for the holonomy symmetries $\Delta^\hbar_L\Psi=0$. Consequently, the anomaly is cancelled at one loop, $\delta^\hbar_L\Gamma=0+{\cal O}(\hbar^2)$, provided that $L^\hbar$ satisfies
\begin{equation}\label{corrected_conditions}
    \hat{\nabla}^\hbar L^\hbar = 0~, \quad i_{L^\hbar}F^\hbar = 0~.
\end{equation}
This scenario is compatible with both the Green--Schwarz mechanism \cite{Green:1984sg} and with the fact that the Killing spinor equations of heterotic supergravity retain
their form up to and including two loops in the sigma model perturbation theory \cite{Bergshoeff:1989de} provided that $H$ is replaced by $H^\hbar$. Assuming that supersymmetry is preserved under quantum corrections, the corrected Killing spinors satisfy $\hat{\nabla}^\hbar\epsilon^\hbar=0$, and therefore, their form bilinears $L^\hbar$ are automatically $\hat{\nabla}^\hbar$-covariantly constant. Furthermore, the corrected gaugino Killing spinor equation implies that $i_{L^\hbar }F^\hbar=0$. The preservation of reduced holonomy under first-order corrections has been established in certain heterotic backgrounds with $SU$ structure; see e.g., \cite{Gillard:2003jh}.

Note that although $H^\hbar$ is no longer closed, $dH^\hbar=-\frac{\hbar}{4\pi}P_4(R,F)$, this correction is necessary to restore the tensorial properties of $H$. Indeed, the anomalous variations of $b$ required to cancel the spacetime frame rotation and gauge anomalies, $\delta_\ell b=\frac{\hbar}{4\pi}Q^1_2(\omega,\ell)$ and $\delta_u b=-\frac{\hbar}{4\pi}Q^1_2(\Omega,u)$, respectively, induce a non-trivial transformation of $H$, which is cancelled by the Chern--Simons contribution to $H^\hbar$.

Having established the conditions for anomaly cancellation, we can now express the Wess--Zumino consistency conditions (\ref{WZ_CC_hol}) in terms of the corrected quantities. Assuming $L^\hbar$ and $M^\hbar$ satisfy (\ref{corrected_conditions}), varying (\ref{corrected_anomaly}) and taking the commutator, one obtains the following expression for the bosonic part \cite{Grimanellis:2023nav}
\begin{equation}\label{WZ_corrected}
    -i\int \dd^2\sigma\dd \theta^+ \left( -2g_{\mu\nu}[\delta^\hbar_L, \delta^\hbar_M]X^\mu \hat{\nabla}^\hbar_= D_+X^\nu + dH^\hbar_{\mu\nu\rho\sigma}\delta^\hbar_LX^\mu \delta^\hbar_MX^\nu D_+X^\rho\partial_=X^\sigma \right) = 0 + {\cal O}(\hbar^2)~.
\end{equation}

The analysis of these conditions depends on the structure of the commutator of holonomy symmetries, which, as we shall see in Section \ref{commutators}, takes the general form 
\begin{equation}\label{G2_algebra}
    [\delta^\hbar_L,\delta^\hbar_M] = \delta^\hbar_N + \delta^\hbar_S + \delta^\hbar_{JP} + \delta^\hbar_{\cal N}~.
\end{equation}
Here, $\delta^\hbar_N$ is generated by a $\hat{\nabla}^\hbar$-covariantly constant form $N$, while $\delta^\hbar_{\cal N}$ and $\delta^\hbar_S$ are given by (\ref{nijen_form_transf}) and (\ref{S_transf}), respectively. The parameters of these transformations are constructed from $a_L$ and $a_M$. Moreover, the transformation $\delta^\hbar_{JP}$ is generated by $\hat{\nabla}^\hbar$-covariantly constant forms collectively denoted by $P$, and its parameter depends on $a_L$, $a_M$ and a set of conserved currents $\cal J$ of the theory, $a_{JP}=a_{JP}(a_L,a_M,{\cal J})$. Hence, (\ref{G2_algebra}) defines a W-algebra, as its structure constants are field-dependent.

Compared with the corresponding algebra for $SU$ holonomy backgrounds \cite{Grimanellis:2023nav}, the algebra (\ref{G2_algebra}) contains an additional transformation $\delta^\hbar_{\cal N}$. The remaining three transformations coincide with those appearing in the $SU$ holonomy commutator and are therefore classified as types I, II, and III, respectively, following the same terminology.

\paragraph{Type I.}

When the commutator closes to a $\delta^\hbar_N$ transformation, only the second term in (\ref{WZ_corrected}) contributes, since $N$ is covariantly constant and $\partial_=a_N=0$. The type I consistency condition then reads
\begin{equation}\label{CC_type_I}
    P_4(R,F)_{\mu\nu[\rho|\sigma|}L^\mu{}_{\cal L}M^\nu{}_{{\cal M}]} = 0~.
\end{equation}
\paragraph{Type II.} When the commutator closes to a $\delta^\hbar_S$ transformation, the first term of (\ref{WZ_corrected}) contributes to the consistency condition, unlike the previous case. The invariance of the action under these transformations relies on the Bianchi identity, which, for non-vanishing $dH$, takes the form
\begin{equation}
    \hat{R}_{[\mu\nu,|\lambda|\rho]} = -\frac{1}{3}\hat{\nabla}_\lambda H_{\mu\nu\rho} - \frac{1}{6}dH_{\lambda\mu\nu\rho}~.
\end{equation}
The resulting consistency condition is 
\begin{equation}\label{CC_type_II}
    \alpha_S\frac{(-1)^q}{3(q+1)} P_4(R,F)_{\sigma[\rho\kappa\lambda}Q_{{\cal Q}]} - a_L a_M P_4(R,F)_{\mu\nu[\rho|\sigma|}L^\mu{}_{\cal L}M^\nu{}_{{\cal M}]} = 0~,
\end{equation}
where $\alpha_S$ is determined by $a_L$ and $a_M$, and the indices are arranged such that $\kappa\lambda {\cal Q} = {\cal LM}$. 

\paragraph{Type III.} Consider now the case where the commutator closes onto a $\delta^\hbar_{JP}$ transformation. Simply promoting the parameter of a symmetry generated by a $\hat{\nabla}$-covariantly constant form $P$ to depend on $\sigma^=$, $\partial_=a_P\neq 0$, does not, in general, preserve the classical action (\ref{het_action}). Nevertheless, since the commutator of two symmetries must itself be a symmetry, the invariance of $\delta^\hbar_{JP}$ can be established by expressing it in terms of $\hat{\nabla}$-covariantly constant forms $L'$ and $M'$ satisfying (\ref{inv_cond}) and 
\begin{equation}\label{JP_split}
    \delta_{JP} = (m'+1)c_{L'}J_{L'}\delta_{M'} + (\ell'+1)c_{M'} J_{M'}\delta_{L'}~.
\end{equation}
Here, $J_{L'}$ and $J_{M'}$ are the conserved currents associated with $L'$ and $M'$, respectively, while the parameters $a_{L'}$ and $a_{M'}$ are determined by $a_L$ and $a_M$ and satisfy the relation $(-1)^{(\ell'+1)(m'+1)}c_{L'}a_{M'} = c_{M'}a_{L'}$. Using this decomposition, the variation of the classical action reduces to a total derivative and vanishes. Repeating the computation in the quantum theory, one finds that the first term of (\ref{WZ_corrected}) likewise makes no contribution, so that the consistency condition reduces to the type I condition (\ref{CC_type_I}). This construction can be generalised to transformations of the form
\begin{equation}
    \delta_{JP} = c_B J_A \delta_B + c_A J_B \delta_A~,
\end{equation}
where $\delta_A$ and $\delta_B$ need not be generated by $\hat{\nabla}$-covariantly constant forms.  

\paragraph{Type IV.} Finally, consider the case where the commutator closes to a $\delta^\hbar_{\cal N}$ transformation, which we call type IV. As we shall see in Section \ref{commutators}, ${\cal N}$ as defined in (\ref{Nijen_form}) is covariantly constant, and the associated transformation has a field-independent parameter. At the quantum level, the $\hat{\nabla}^\hbar$-covariant constancy of ${\cal N}^\hbar$ is also preserved, so the corresponding consistency condition reduces to that of type I. Extending this investigation to an ${\cal N}$ with non-zero components along the fibres would require additional care, since the relations (\ref{PQ_eqn}) hold only in the space where the invariant forms of the background are defined, here the base space. Nevertheless, this case is beyond the scope of our analysis, as the only type IV symmetry that arises in our investigation is horizontal, $i_K{\cal N}=0$. 

Although we have considered the consistency conditions associated with each type of transformation separately, the derivation of the corresponding conditions in Section \ref{sec_an_cc} will account for the sum of all transformations appearing on the right-hand side of each commutator.

\section{W-algebra and anomalies for \texorpdfstring{$G_2$}{G₂} backgrounds}\label{section_3}

\subsection{The \texorpdfstring{$G_2$}{G₂} geometry}\label{section_the_G2_geometry}

Heterotic backgrounds with compact holonomy can be locally described as fibrations of $G$ over a base space $B$, with the Killing vectors generating the Lie algebra of $G$ acting on spacetime $M$. Thus, these backgrounds can be modelled locally as principal $G$-bundles $P(G,B,\pi)$ with a principal bundle connection $e^a$. In particular, for heterotic backgrounds with $G_2$ holonomy, the base space is a seven-dimensional Riemannian manifold $N^7$, while the fibres are diffeomorphic to the three-dimensional Lie group $G$. The base space $N^7$ is conformally balanced, and it is equipped with a metric $\tilde{g}$ and torsion $\tilde{H}$. For further details on the geometry of heterotic backgrounds and the $G_2$ structures, see \cite{Gran:2005wf,Gran:2007fu,Gran:2018ijr,Friedrich:2001nh, Friedrich:2001yp}.

To describe the geometry of these backgrounds, we introduce a pseudo-orthonormal coframe $e^A=(e^a,e^i)$, with $e^a$ orthogonal to $e^i$, where $a$ and $i$ are fibre and base space indices, respectively. In that coframe, the metric and the 3-form field strength are written as
\begin{align}\label{metric_flux_definition}
    ds^2 &= \eta_{ab}e^a e^b + d\tilde{s}^2~, \nn
    H&= \frac{1}{3}\eta_{ab} e^a\wedge d e^b + \frac{2}{3}\eta_{ab} e^a\wedge{\cal F}^b + \tilde{H}~ \nn
    &= \frac{1}{3!}H_{abc}e^{abc} + \frac{1}{2}H_{aij}e^{aij} + \frac{1}{3!}H_{ijk}e^{ijk}~,
\end{align}
where $e^{ijk}$ is a shorthand notation for $e^i\wedge e^j \wedge e^k$, $d\tilde{s}^2=\delta_{ij}e^i e^j$ is the metric on the base space, and the components of $\tilde{H}$ coincide with the purely horizontal components of $H$, namely $\tilde{H}_{ijk}=H_{ijk}$. Moreover,
\begin{equation}\label{PB_vcurvature}
    {\cal F}^a \coloneqq \frac{1}{2}H_{aij} e^{ij} = d e^a - \frac{1}{2}H^a{}_{bc} e^{bc}~,
\end{equation}
is the principal bundle curvature, where $H$ is taken to be closed $dH = \eta_{ab}{\cal F}^a\wedge{\cal F}^b + d\tilde{H}=0$. In general, as we saw, the Green--Schwarz mechanism \cite{Green:1984sg} assigns an anomalous contribution to $dH \propto \text{tr}R^2 - \text{tr}F^2$, where $R$ is the curvature of a connection on the tangent bundle, whose choice is renormalisation-scheme dependent and can be left arbitrary for now.

\paragraph{$\hat{\nabla}$-covariantly constant forms.}

There are two types of $\hat{\nabla}$-covariantly constant forms in a heterotic background. The first type is a set of three 1-forms $e^a$, and the second type is a 3-form $\varphi\in\Lambda^3(N^7)$ that is responsible for the reduction of the holonomy to $G_2$, and its dual $\star_7\varphi\equiv \psi\in\Lambda^4(N^7)$. 

Regarding the set of the 1-forms $e^a$, one of them is timelike, and the other two are spacelike, as in all cases with a compact holonomy group. From their $\hat{\nabla}$-covariant constancy it follows \cite{Gran:2005wf}
\begin{equation}
    {\cal L}_{e_a}g=0~,
\end{equation}
which implies that the vector fields $K_a=e_a$, dual to $e^a$, are Killing. These are constructed as Killing spinor bilinears and, in general, span only a subalgebra of the full spacetime isometry algebra. Thus, $G$ is a subgroup of the full isometry group. It follows from the Killing spinor equations that the Killing vector fields $K_a$ satisfy the algebra
\begin{equation}\label{KV_algebra}
    [K_a,K_b]=-H^A{}_{ab}K_A~.
\end{equation}

Denoting by $\mathfrak{g}=\text{span}(K_a)$ the Lie algebra spanned by $K_a$, the relation $[\mathfrak{g},\mathfrak{g}]\subseteq\mathfrak{g}$ implies that the components $H_{abi}$ vanish. Thus, the structure constants of the Lie algebra are identified with $H_{abc}$ and are generally not required to vanish from the Killing spinor equations (\ref{het_KSE_1}) and (\ref{het_KSE_2}), leaving two possible options. For $H_{abc}$ vanishing, $\mathfrak{g}$ is isomorphic to $\mathfrak{u}(1)\oplus \mathfrak{u}(1)\oplus \mathfrak{u}(1)$, otherwise it is isomorphic to $\mathfrak{sl}(2,\mathbb{R})$. Moreover, it follows that $i_{e_a}\tilde{H}=i_{e_a}{\cal F}=0$.

On the other hand, 
\begin{equation}
    \varphi = \frac{1}{3!}\varphi_{ijk}e^{ijk}~,
\end{equation}
is the $G_2$-invariant form and lives in the base space $N^7$, as it is horizontal and invariant under the action of the Killing vector fields
\begin{equation}
    i_{e_a}\varphi=0~, \quad {\cal L}_{e_a}\varphi=0~,
\end{equation}
that translates into $\hat{\nabla}\varphi = 0$ and hence $\hat{\tilde{\nabla}}\varphi =0$, implying that the Riemannian manifold $N^7$ with metric connection $\hat{\tilde{\nabla}}$, admits a $G_2$-structure as $\text{hol}(\hat{\tilde{\nabla}})\subseteq G_2$. Similarly, for the 4-form $\psi$. 

As the dilatino Killing spinor equation (\ref{het_KSE_2}) implies that $(i_{e_a}H)_{ij}$ take values in $\mathfrak{g}_2$, the principal bundle curvature ${\cal F}$ (\ref{PB_vcurvature}) is a $G_2$-instanton on the base space
\begin{equation}
    {\cal F}_{ij} = \frac{1}{2}\psi_{ij}{}^{kl}{\cal F}_{kl}~,
\end{equation}
implying the condition
\begin{equation}\label{G2_instanton_1}
    H_{aij}\epsilon^{ij}{}_{lmnpq} = 20\varphi_{[lmn}H_{pq]a}~,
\end{equation}
which in turn, contracted by $\psi^{lmnp}$ and using the identities in Appendix \ref{subsection_identities} translates to
\begin{equation}\label{G2_instanton_2}
    \varphi_{ijk}H^{ij}{}_a = 0~.
\end{equation}

Both $\varphi$ and $\psi$ are Killing spinor bilinears as they are constructed from the Killing spinors of $G_2$ backgrounds. Choosing a coframe $(e^2,e^3,e^4,e^6,e^7,e^8,e^9)$ for the base space, $\varphi$ is defined as
    \begin{align}
        \varphi = \text{Re}\hat{\chi} + e^6 \wedge\hat{\omega}~,
    \end{align}
with 
\begin{equation}
        \hat{\omega} = -e^2\wedge e^7 - e^3\wedge e^8 - e^4\wedge e^9~, \quad \hat{\chi} = (e^2+ie^7)\wedge(e^3+ie^8)\wedge(e^4+ie^9)~.
\end{equation}

For the definition of $\psi=\star_7\varphi$, the Hodge dual in the base space is taken with respect to the volume form $ e^2\wedge e^3\wedge e^4\wedge e^6\wedge e^7\wedge e^8\wedge e^9$. With these conventions, the above forms can be expressed as
\begin{align}\label{phi_psi_frame}
    \varphi &= e^{234} - e^{289} + e^{379} - e^{478} - e^{627} - e^{638} - e^{649}~, \nn
        \psi &= e^{6789} - e^{3467} + e^{2468} - e^{2369} + e^{3489} + e^{2479} + e^{2378}~.
\end{align}

\subsection{Commutators of holonomy symmetries}\label{commutators}

Having established the relevant geometric properties, we may now proceed with the computation of the commutators. The symmetry transformations for the action (\ref{het_action}) associated with the spacetime isometries and the invariant forms of $G_2$ are given by
\begin{equation}
    \delta_K X^\mu = a_K^a e^\mu_a~, \quad \delta_\varphi X^\mu = a_\varphi \varphi^\mu{}_{\nu\rho} D_+X^{\nu\rho}~, \quad 
    \delta_\psi X^\mu = a_\psi \psi^\mu{}_{\nu\rho\sigma} D_+ X^{\nu\rho\sigma}~.  
\end{equation}
For the closure of the algebra, one additionally has to consider the transformation 
\begin{equation}\label{nijen_symmetry}
    \delta_{\cal N}X^\mu = a_{\cal N} {\cal N}(\psi,\psi)^\mu{}_{\nu_1\dots\nu_6}D_+X^{\nu_1\dots\nu_6}~,
\end{equation}
where $\cal{N}(\psi,\psi)$ is the Nijenhuis form (\ref{Nijen_form}) for $L=M=\psi$, restricted along the orthogonal directions to those of the fibres. In terms of the pseudo-orthonormal frame $e^A$, one has $\delta X^A = e^A_\mu \delta X^\mu$, $D_+X^A = e^A_\mu D_+X^\mu$, and $\partial_=X^A = e^A_\mu \partial_= X^\mu$, and the above transformations can be written as 
\begin{gather}
     \delta_K X^a = a^a_K ~, \quad \delta_\varphi X^i = a_\varphi \varphi^i{}_{jk} D_+X^{jk}~, \nn 
    \delta_\psi X^i = a_\psi \psi^i{}_{jkl} D_+ X^{jkl}~, \quad \delta_{\cal N}X^i = a_{\cal N} {\cal N}(\psi,\psi)^i{}_{i_1\dots i_6}D_+X^{i_1\dots i_6}~.
\end{gather}
The proof of (\ref{nijen_symmetry}) being a classical symmetry follows from the covariant constancy of $\cal N(\psi,\psi)$. As has been shown in (\ref{modified_nijenhuis}), the Nijenhuis form can be expressed as 
\begin{equation}\label{nijenhuis_form_epsilon}
    {\cal N(\psi,\psi)}_{\ell_1\dots\ell_7} = \frac{1}{10}\epsilon_{\ell_1\dots \ell_7} \varphi_{ijk} H^{ijk}~.
\end{equation}
Moreover, the dilatino Killing spinor equation implies (see e.g., \cite{Gran:2018ijr})
\begin{equation}\label{dilatino_kse}
    \epsilon^{abc}H_{abc} + H_{ijk}\varphi^{ijk} = 0~.
\end{equation}
From the Lie algebra (\ref{KV_algebra}) and the $\hat{\nabla}$-covariant constancy of the Killing vectors, it follows that the structure constants $H_{abc}$ are also annihilated by $\hat{\nabla}$. Then from (\ref{nijenhuis_form_epsilon}) and (\ref{dilatino_kse}), ${\cal N}(\psi,\psi)$ is $\hat{\nabla}$-covariantly constant and therefore (\ref{nijen_symmetry}) is an individual symmetry of the action (\ref{het_action}). The same argument holds at the quantum level, assuming that the corrected Killing spinor bilinears remain $\hat{\nabla}^\hbar$-covariantly constant. Note that this type of symmetry did not appear in the $SU$ holonomy cases in \cite{Grimanellis:2023nav}.

Another symmetry transformation needed for the closure of the algebra is the one associated with the quadratic Casimir operator of the Lie algebra $\mathfrak{g}$ of the isometries 
\begin{equation}
    C = \eta_{ab}e^a_\mu e^b_\nu D_+X^\mu\hat{\nabla}_+D_+X^\nu~,
\end{equation}
and is given by
\begin{equation}
    \delta_C X^a = \alpha_C \hat{\nabla}_+ D_+X^a + \hat{\nabla}_+(\alpha_CD_+ X^a)~,
\end{equation}
with $\delta_CX^i=\Delta_C\Psi_-^\ma=0$. Using the algebra of the Killing vectors (\ref{KV_algebra}), one finds

\begin{equation}\label{commutator_KK}
    [\delta_K,\delta_K']X^\mu = a^a_K (a'_K)^b [e_a,e_b]^\mu = -a^a_K (a'_K)^b H_{ab}{}^c e_c^\mu = \delta''_K X^\mu~, 
\end{equation}
with $(a_K'')^c=-a^a_K(a'_K)^bH_{ab}{}^c$, and $\Delta_{KK}\Psi^\ma_-=\Delta_K\Psi^\ma_-=0$. Moreover, the commutators of $\delta_K$ with $\delta_\varphi$ and $\delta_\psi$ are
\begin{align}
    [\delta_K, \delta_\varphi]X^\mu &= a_\varphi a^a_K ({\cal L}_{e_a}\varphi)^\mu{}_{\nu\rho} D_+X^{\nu\rho} = 0~, \label{commutator_K_phi} \\ 
    [\delta_K, \delta_\psi]X^\mu &= a_\psi a^a_K ({\cal L}_{e_a}\psi)^\mu{}_{\nu\rho\sigma}D_+X^{\nu\rho\sigma}=0~, \label{commutator_K_psi}
\end{align}
as $\varphi$ and $\psi$ are invariant under isometries. The commutator acting on $\Psi_-$ given by (\ref{psi_commutator}) also vanishes as a consequence of (\ref{inv_cond}).

\paragraph{$\varphi-\varphi$ commutator.} With a straightforward computation, using (\ref{LM_term_1})-(\ref{LM_term_3}), the only non-vanishing term is
\begin{equation}
    [\delta_\varphi, \delta'_\varphi]X^i = \delta_{\varphi\varphi}^{(2)}X^i
~,\end{equation}
with
\begin{equation}
    \delta^{(2)}_{\varphi\varphi'}X^i = 2 (a_\varphi'D_+a_\varphi-a_\varphi D_+ a'_\varphi)\varphi_{mi[j}\varphi^m{}_{kl]}D_+X^{jkl}~.
\end{equation}
As the relations (\ref{PQ_eqn}) give $P=\psi$ and $Q=0$, there are no $\delta_S$ or $\delta_{\cal N}$ contributions on the right-hand side of the commutator. Thus, using the identities in Appendix \ref{subsection_identities}, we obtain

\begin{equation}
    \delta^{(2)}_{\varphi\varphi'} X^i = -2 (a_\varphi'D_+a_\varphi-a_\varphi D_+ a'_\varphi) \psi^i{}_{jkl}D_+X^{jkl}~,
\end{equation}
and therefore
\begin{equation}\label{commutator_phi_phi}
    [\delta_\varphi, \delta'_\varphi] = \delta_{\psi}~,
\end{equation}
with $a_\psi = -2 (a_\varphi'D_+a_\varphi-a_\varphi D_+ a'_\varphi)$, which agrees with \cite{Howe:1991ic}. The commutator of two $\delta_\varphi$ transformations on $\Psi_-$ is given from (\ref{psi_commutator}) with $\Delta_{\varphi\varphi}\Psi^\ma_-=\Delta_\psi\Psi^\ma_-=0$, which is consistent with $i_\varphi F=0$ required for $\delta_\varphi$ to be a symmetry of the action (\ref{het_action}).

\paragraph{$\varphi-\psi$ commutator.}

To proceed with the commutator of the symmetries generated by $\varphi$ and $\psi$, the non-vanishing terms of (\ref{LM_term_1})-(\ref{LM_term_3}) are
\begin{align}
    \delta^{(1)}_{\varphi\psi}X_a &= a_\psi a_\varphi N(\varphi,\psi)_{a,lmnpq} D_+X^{lmnpq}~, \label{delta_1a} \\
    \delta^{(1)}_{\varphi\psi}X_i &= a_\psi a_\varphi N(\varphi,\psi)_{i,lmnpq} D_+X^{lmnpq}  + 5 a_\psi a_\varphi N(\varphi,\psi)_{i,almnp}D_+X^{almnp}~, \label{delta_1i} \\ 
    \delta^{(2)}_{\varphi\psi}X_i &= \left( -3a_\psi D_+ a_\varphi \varphi_{j[lm}\psi^j{}_{|i|np]} + 2a_\varphi D_+a_\psi \varphi_{ij[l}\psi^j{}_{mnp]} \right) D_+X^{lmnp}~, \label{delta_2i} \\ 
    \delta^{(3)}_{\varphi\psi}X_i &= -6ia_\psi a_\varphi \left( \varphi_{ji[l} \psi^j{}_{|k|mn]} + \varphi_{jk[l}\psi^j{}_{|i|mn]} \right) \partial_{\pp} X^kD_+X^{lmn}~, \label{delta_3i}
\end{align}
where, from the definition of the Nijenhuis tensor (\ref{Nijenhuis_tensor}), one obtains
\begin{align}
    N(\varphi,\psi)_{a,lmnpq}
    &= - H_{ajk} \varphi^j{}_{[lm} \psi^k{}_{npq]}~, \label{Nijen_1} \\
    N(\varphi,\psi)_{i,lmnpq}
    &= - H_{ijk}\varphi^j{}_{[lm} \psi^k{}_{npq]}
    - 3H_{jk[l} \varphi^j{}_{mn} \psi_{pq]i}{}^k
    - 2H_{jk[l}\varphi_{m|i|}{}^j \psi^k{}_{npq]}
    \label{Nijen_2} \nn
    &\hphantom{{}={}}\!+ 3H_{j[lm}\varphi_{|ki|n}\psi^{jk}{}_{pq]}
    - 3H^j{}_{[lm}\varphi_{|jk|n}\psi_{pq]i}{}^k ~, \\
    N(\varphi,\psi)_{i,almnp}
    &= \frac{1}{5}\left(
    -3H_{ajk}\varphi^j{}_{[lm}\psi_{np]i}{}^k
    - 2H_{ajk}\varphi_{[l|i|}{}^j\psi^k{}_{mnp]}
    \right. \label{Nijen_3} \nn
    &\hphantom{{}={}\frac{1}{5}(}
    \left. + 6H_{ja[l}\varphi_{|ki|m}\psi^{jk}{}_{np]}
    - 6H^j{}_{a[l}\varphi_{|jk|m}\psi_{np]i}{}^k
    \right)~.
\end{align}

To compute these components, we use the fact that the Hodge star operation is invertible, with $(\star_7)^2=1$ for the base space $B=N^7$. Then for the first Nijenhuis tensor, we obtain
\begin{align}
    (\star_7 N(\varphi,\psi)_a)_{lm} &= \frac{1}{5!}N(\varphi,\psi)_{a,i_1\dots i_5}\epsilon^{i_1\dots i_5}{}_{lm} \nn
    &=\frac{1}{5\cdot12}\left( 2H_{an[l}\psi_{m]}{}^{pqr}\psi^n{}_{pqr} + 3H^{p}{}_{an}\psi^{qr}{}_{lm}\psi^n{}_{qpr} \right)~,
\end{align}
and employing the identities in Appendix \ref{subsection_identities}, it becomes
\begin{align}
    (\star_7 N(\varphi,\psi)_a)_{lm} = -\frac{2}{5}H_{alm} - \frac{1}{10}H_{ajk}\psi^{jk}{}_{lm}~.
\end{align}
Dualising once again
\begin{align}
    N(\varphi,\psi)_{a,lmnpq} = -\frac{1}{5}H_{ajk}\epsilon^{jk}{}_{lmnpq} - 2H_{a[lm}\varphi_{npq]}~,
\end{align}
and using the $G_2$-instanton condition (\ref{G2_instanton_1}) we find
\begin{equation}
    N(\varphi,\psi)_{a,lmnpq} = -6H_{a[lm}\varphi_{npq]}~.
\end{equation}
Similarly, the rest of the components read as follows
\begin{align}\label{nijen_results}
    N(\varphi,\psi)_{i,almnp} &= 0~, \nn
    N(\varphi,\psi)_{i,lmnpq} &= - 3H_{[lmn}\varphi_{pq]i} - 9g_{i[l}H^j{}_{mn}\varphi_{pq]j}~.
\end{align}
    
To express the right-hand side of
\begin{gather}
    [\delta_\varphi, \delta_\psi]X_a = \delta^{(1)}_{\varphi\psi}X_a~, \\
    [\delta_\varphi, \delta_\psi]X_i = \delta^{(1)}_{\varphi\psi}X_i + \delta^{(2)}_{\varphi\psi} X_i + \delta^{(3)}_{\varphi\psi}X_i~,
\end{gather}
as a sum of individual symmetries, we further need the symmetry (\ref{S_transf}) $S^\mu{}_{\nu\rho\sigma\tau} = \delta^\mu{}_{[\nu}Q_{\rho\sigma\tau]}$. From the relations (\ref{PQ_eqn}) for $L=\varphi$ and $M=\psi$, one derives
\begin{equation}
    P = 0~, \quad Q = -2\varphi~,
\end{equation}
and the corresponding transformation along the base space directions is 
\begin{align}\label{S_symmetry_P}
    \delta_SX_i = -\delta'_\varphi X_i & - \frac{i}{2}\alpha_S\left( 2g_{ij}\varphi_{klm} - 3g_{j[k}\varphi_{lm]i} - 3g_{i[k}\varphi_{lm]j} \right) \partial_{\pp}X^j D_+X^{klm} \nn 
    &-\frac{1}{4}\alpha_S \left( -H_{[jkl}\varphi_{mn]i} - 3g_{i[j}H^p{}_{kl}\varphi_{mn]p} \right)D_+X^{jklmn} \nn 
    &+\frac{1}{2}(D_+\alpha_S)g_{i[j}\varphi_{klm]}D_+X^{jklm}+\frac{3}{2}\alpha_S H^p{}_{a[j}g_{|i|k}\varphi_{lm]p}D_+X^{ajklm}~,
\end{align}
with $a'_\varphi = -\frac{3}{2}\alpha_S C$, and $\hat{\nabla}_+$ anticommuting with $\alpha_S$. The last term drops as $\varphi$ is invariant under the action of the Killing vector fields, $({\cal L}_{e_a}\varphi)_{ijk} = -3H_a{}^p{}_{[i}\varphi_{jk]p}=0$. Therefore, from (\ref{nijen_results}) and using the identities in Appendix \ref{subsection_identities} for $\delta^{(2)}_{\varphi\psi}$ and $\delta^{(3)}_{\varphi\psi}$  given by (\ref{delta_2i}) and (\ref{delta_3i}), respectively, we can write
\begin{equation}
    \delta_SX_i = -\delta'_\varphi X_i +  \delta^{(1)}_{\varphi\psi}X_i + \delta^{(2)}_{\varphi\psi}X_i + \delta^{(3)}_{\varphi\psi}X_i ~,
\end{equation}
for $\alpha_S = -12a_\psi a_\varphi$. With a similar manipulation, along the fibre directions the $S$-transformation can be expressed as
\begin{equation}\label{delta_S_Xa}
    \delta_S  X_a =  -\delta_CX_a - \delta_K X_a - 6a_\psi a_\varphi H_{a[lm}\varphi_{npq]}D_+X^{lmnpq}~,
\end{equation}
with $\alpha_C=\frac{1}{2}\alpha_S J_\varphi$, and $a^a_K = -\frac{1}{2}\alpha_S H^a{}_{bc}J^b_K J^c_K J_\varphi$, where $J_\varphi=\varphi_{ijk}D_+X^{ijk}$ and $J^a_K=D_+X^a$ the conserved currents for $\varphi$ and the isometries $e^a_\mu$, respectively. Therefore, the $\varphi-\psi$ commutator along all spacetime directions can be written as
\begin{equation}\label{commutator_phi_psi}
    [\delta_\varphi, \delta_\psi] = \delta_S + \delta'_\varphi + \delta_C + \delta_K~,
\end{equation}
with the transformation parameters given by
\begin{equation}
    a'_\varphi = 18 a_\psi a_\varphi C~, \quad \alpha_C = -6a_\psi a_\varphi J_\varphi~, \quad a^a_K = 6a_\psi a_\varphi H^a{}_{bc}J^b_K J^c_K J_\varphi~.
\end{equation} 
Note that ${\cal N}(\varphi,\psi)$ vanishes by virtue of the identity
\begin{equation}\label{H-phi_identity}
    H_{pq[i}\varphi^p{}_{jk}\psi^q{}_{lmn]} = 2H_{[ijk}\varphi_{lmn]}~.
\end{equation}
Moreover, the commutator of the transformations $\delta_\varphi$ and $\delta_\psi$ on $\Psi_-$ is given by (\ref{psi_commutator}) with 
\begin{equation}
    \Delta_{\varphi\psi}\Psi_-^\ma = \Delta_S\Psi^\ma_- = 6a_\psi a_\varphi \varphi_{[\mu\nu\rho} F_{\sigma\kappa]}{}^\ma{}_\mb \Psi_-^\mb D_+X^{\mu\nu\rho\sigma\kappa}~.
\end{equation}

\paragraph{$\psi-\psi$ commutator.}
When it comes to the commutator generated by two symmetries $\psi$, we find that
\begin{align}
    \delta^{(1)}_{\psi\psi'}X_a &=a'_\psi a_\psi N(\psi,\psi)_{a,lmnpqr}D_+X^{lmnpqr}~,  \\ 
    \delta^{(1)}_{\psi\psi'}X_i &= a'_\psi a_\psi N(\psi,\psi)_{i,lmnpqr}D_+X^{lmnpqr} + 6a'_\psi a_\psi N(\psi,\psi)_{i,almnpq}D_+X^{almnpq}~, \\  
    \delta^{(2)}_{\psi\psi'}X_i &=  3\left( a'_\psi D_+a_\psi - a_\psi D_+ a'_\psi \right) \psi_{ij[lm} \psi^j{}_{npq]}D_+X^{lmnpq}~, \\
    \delta^{(3)}_{\psi\psi'}X_i &= -18ia'_\psi a_\psi \psi_{ij[lm} \psi^j{}_{np]k} \partial_{\pp} X^k D_+ X^{lmnp}~,
\end{align}
with the components of the Nijenhuis tensor given by (\ref{Nijenhuis_tensor})
\begin{align}
    N(\psi,\psi)_{a,lmnpqr} &= - H_{ajk} \psi^j{}_{[lmn}\psi^k{}_{pqr]}~, \label{Npsipsi_1} \\
    N(\psi,\psi)_{i,lmnpqr} &=  -H_{ijk} \psi^j{}_{[lmn} \psi^k{}_{pqr]} + 6\psi^j{}_{[lmn}H_{|j|p}{}^k \psi_{qr]ik} + 9H^j{}_{[lm} \psi_{|ki|np} \psi^k{}_{|j|qr]}~, \label{Npsipsi_2}\\
    N(\psi,\psi)_{i,almnpq} &= -6H^k{}_{ja} \psi^j{}_{[lmn} \psi_{pq]ik} + 18H^k{}_{a[l} \psi_{|ji|mn} \psi^j{}_{pq]k} ~. \label{Npsipsi_3}
\end{align}
    
For their computation, we shall take again twice the Hodge dual with respect to the skew-symmetric block of base space indices. Starting from (\ref{Npsipsi_1}), its Hodge dual reads
\begin{equation}
    (\star_7 N(\psi,\psi)_a)_l =\frac{1}{10} H_{amn}\varphi^{mn}{}_l ~,
\end{equation}
which vanishes from the $G_2$-instanton condition (\ref{G2_instanton_2}), and therefore, 
\begin{equation}\label{Hpsi_condition}
    N(\psi,\psi)_{a,ijklmn} =3 H_{a[ij}\psi_{klmn]}=0~.
\end{equation}
Similarly, the Hodge dual of (\ref{Npsipsi_2}) reads
\begin{equation}
    (\star_7 N(\psi,\psi)_i)_l = \frac{1}{5}H_{ijk}\varphi^{jk}{}_l - \frac{1}{10}H_{jkl}\varphi^{jk}{}_i~,
\end{equation}
and thus it follows from the identities in Appendix \ref{subsection_identities} that
\begin{align}
    N(\psi,\psi)_{i,lmnpqr} &= 3 H_{i[lm}\psi_{npqr]} - 12g_{i[l}H^s{}_{mn}\psi_{pqr]s}~.
\end{align}
    
For the calculation of the last term (\ref{Npsipsi_3}) one can readily verify that
\begin{equation}
    \left(\star_7 N(\psi,\psi)_{i,a}\right)_{lm} = - \frac{1}{5}H^{jk}{}_a g_{i[l}\varphi_{m]jk} -\frac{1}{5}{\cal L}_a\varphi_{ilm} ~,
\end{equation}
which vanishes due to the $G_2$-instanton condition (\ref{G2_instanton_2}) and the invariance of $\varphi$ under isometries, giving $N(\psi,\psi)_{i,almnpq}=0$. To express the $\psi-\psi$ commutator as a sum of individual symmetries, we shall, moreover, need the ${\cal N}$ and $S$-type of symmetries (\ref{nijen_form_transf}) and (\ref{S_transf}), respectively. It then follows from the relations (\ref{PQ_eqn}) for $L=M=\psi$ that
\begin{equation}
    P = 0~, \quad Q = -\frac{4}{3}\psi~.
\end{equation}
Therefore, along the base space directions and using the condition (\ref{Hpsi_condition}), the $S$-transformation reads
\begin{align}
    \delta_SX_i = -\delta''_\psi X_i
    &- \frac{4}{15}\alpha_S \left( 2ig_{ik} \psi_{lmnp}
    + 4ig_{k[l}\psi_{mnp]i}
    + 4ig_{i[l}\psi_{mnp]k}\right)
    \partial_{\pp}X^k D_+X^{lmnp} \nn
    &- \frac{8}{15}\alpha_S
    \left(\frac{1}{3}H_{[lmn} \psi_{pqr]i}
    + g_{i[l}\psi_{mnp|j|} H^j{}_{qr]} \right)
    D_+X^{lmnpqr} \nn
    &- \frac{4}{15}(D_+\alpha_S)g_{i[l}\psi_{mnpq]}D_+X^{lmnpq}~,
\end{align}
with
\begin{align}
    a''_\psi &= -\frac{16}{15}\alpha_S C + \frac{16}{45}\alpha_SH_{abc}J^a_KJ^b_K J^c_K~, 
\end{align}
and similarly, along the fibre directions we get
\begin{equation}
    \delta_S X_a = -\delta_CX_a - \delta_KX_a + \frac{4}{15}\alpha_S J_\psi H_{alm}D_+X^{lm}~, 
\end{equation}
where $J_\psi = \psi_{ijkl}D_+X^{ijkl}$ is the conserved current of $\psi$, and the corresponding parameters read $\alpha_C = \frac{4}{15}\alpha_S J_\psi$ and $a^a_K = -\frac{4}{15}\alpha_S H^a{}_{bc}J^b_KJ^c_KJ_\psi$. In addition, the transformation $\delta_{\cal N}$ given by (\ref{nijen_form_transf}) and generated by the Nijenhuis form ${\cal N}(\psi,\psi)$ reads
\begin{align}
    {\cal N}(\psi,\psi)_{ijklmns} = -7\left( H_{pq[i}\psi^p{}_{jkl}\psi^q{}_{mns]} + 2H_{[ijk}\psi_{lmns]} \right)~.
\end{align}
Using the identities in Appendix \ref{subsection_identities}, its Hodge dual is a scalar that takes the form
\begin{equation}\label{modified_nijenhuis}
    \star_7{\cal N}(\psi,\psi) = \frac{1}{10}H_{ijk}\varphi^{ijk}~.
\end{equation}
Dualising again, one arrives at the final expression
\begin{equation}
    {\cal N}(\psi,\psi)_{ijklmns} = 12\psi_{i[jkl}H_{mns]} + 9\psi_{[jklm}H_{ns]i}~.
\end{equation}
Therefore, the $\psi-\psi$ commutator for all spacetime directions reads
\begin{equation}\label{commutator_psi_psi}
    [\delta_\psi, \delta'_\psi] = \delta_S + \delta_{\cal N} + \delta''_\psi + \delta_C +\delta_K~,
\end{equation}
with the parameters 
\begin{gather}
    \alpha_S = \frac{45}{2}a'_\psi a_\psi ~, \quad  a_{\cal N} = \frac{1}{3}a'_\psi a_\psi ~, \quad  a''_\psi = -24a'_\psi a_\psi C + 8a'_\psi a_\psi H_{abc}J^a_K J^b_K J^c_K~, \nn 
    \alpha_C = 6a'_\psi a_\psi J_\psi ~, \quad a^a_K = -6a'_\psi a_\psi H^a{}_{bc}J^b_K J^c_K J_\psi ~.
\end{gather}
Note that some of these depend on the conserved currents of the theory, which is a characteristic of a $W$-algebra. One can also see that, although they are field-dependent, the transformations can be rearranged in a way such that the action remains invariant, as discussed in Section \ref{section_anomalies}. Furthermore, the commutator of two $\delta_\psi$ transformations acting on $\Psi_-$ is given by (\ref{psi_commutator}) with 
\begin{equation}
    \Delta_{\psi\psi}\Psi_-^\ma = \Delta_S\Psi^\ma_- = 6 a'_\psi a_\psi  \psi_{[\mu\nu\rho\sigma}F_{\kappa\lambda]}{}^{\ma}{}_{\mb}\Psi^{\mb}_- D_+X^{\mu\nu\rho\sigma\kappa\lambda}~.
\end{equation}
    
Collecting the results (\ref{commutator_phi_phi}), (\ref{commutator_phi_psi}), and (\ref{commutator_psi_psi}), the commutators of the holonomy symmetries can be summarised as 
\begin{align}\label{all_commutators}
    {}[\delta_\varphi, \delta'_\varphi] &= \delta_{\psi}~, \nn
    {}[\delta_{\varphi}, \delta_{\psi}] &= \delta_S + \delta_{\varphi}' + \delta_C + \delta_K~, \nn
    {}[\delta_\psi, \delta_{\psi}'] &= \delta_S + \delta_{\cal N} + \delta_{\psi}'' + \delta_C +\delta_K~,
\end{align}
and therefore, they close according to the general expression 
\begin{equation}
    [\delta_L, \delta_M] = \delta_N + \delta_S + \delta_{JP} + \delta_{\cal N}~.
\end{equation}

The presence of field-dependent structure constants indicates that the above expression corresponds to a $W$-algebra. The commutators (\ref{all_commutators}) illustrate the non-trivial role of the fibre directions in the structure of the resulting algebra. Although the $G_2$-invariant forms $\varphi$ and $\psi$ are horizontal, their commutators contain contributions that extend along the fibre directions and must be included to express the algebra as a sum of individual symmetries on the full ten-dimensional spacetime. This distinguishes the resulting algebra from that of sigma models with seven-dimensional $G_2$ target spaces, where such contributions are absent.

Compared to the algebra obtained in the cases of $SU(2)$ and $SU(3)$ holonomy backgrounds, the major difference lies in the symmetry generated by the Nijenhuis form, which is absent in \cite{Grimanellis:2023nav}, although non-vanishing Nijenhuis forms may also arise \cite{Howe:2010az}. However, in the special case where $G$ is abelian, it follows from (\ref{dilatino_kse}) and (\ref{modified_nijenhuis}) that the Nijenhuis symmetry vanishes, leaving an algebra with the same general structure as that derived for heterotic backgrounds with $SU$ holonomy.

\subsection{Anomaly consistency conditions}\label{sec_an_cc}

It remains to examine the Wess--Zumino consistency conditions for each commutator obtained in Section \ref{commutators}. As discussed in Section \ref{section_anomalies}, these conditions depend on the symmetry transformations appearing on the right-hand side of the commutators. In particular, the type I, III and IV transformations alone lead to the same consistency condition (\ref{CC_type_I}), whereas the presence of type II transformations gives rise to an additional contribution as expressed in (\ref{CC_type_II}), which depends on $Q$, defined in (\ref{PQ_eqn}). We shall therefore verify the conditions by identifying the relevant symmetry types in each commutator.

To begin with, the commutators of $\delta_K$ with $\delta_K$, $\delta_\varphi$ and $\delta_\psi$, given by (\ref{commutator_KK})--(\ref{commutator_K_psi}), respectively, either vanish or close onto a type I symmetry. The corresponding consistency condition follows from (\ref{CC_type_I}) and reads

\begin{equation}\label{CC_K_all}
    P_4(R,F)_{a\nu[\rho|\sigma|}M^\nu{}_{{\cal M}]}=0~,
\end{equation}
where $M=K,\varphi,\psi$. Although the connection appearing in the bosonic part of the action is $\hat\nabla$, with torsion $+H$, the appropriate connection for the anomaly is $\check\nabla$, with torsion $-H$ and curvature $\check R$; see \cite{Hull:1986xn}. We therefore express the frame rotation (\ref{frame_anomaly}) and the holonomy symmetry (\ref{hol_anomaly}) anomalies in terms of $\check\omega$. Setting $dH=0$ in (\ref{R_ind_exch}), the curvatures of the two torsionful connections satisfy 

\begin{equation}
    \hat R_{\mu\nu,\rho\sigma}
    =\check R_{\rho\sigma,\mu\nu}~.
\end{equation}

Since the holonomy algebra of $\hat\nabla$ is contained in $\mathfrak{g}_2$, its curvature annihilates the $\hat\nabla$-parallel Killing vectors $K$. The above identity then implies $i_K\check R=0$, and consequently $i_KP_4(\check R)=0$. Furthermore, as the gaugino Killing spinor equation (\ref{het_KSE_1}) implies that the gauge curvature $F$ takes values in $\mathfrak{g}_2$, it follows that $i_KF=0$, and hence $i_KP_4(F)=0$. Together, these give $i_KP_4(\check R,F)=0$, so that the consistency condition (\ref{CC_K_all}) is satisfied for all three commutators involving $\delta_K$.

We next consider the commutator (\ref{commutator_phi_phi}) of two transformations generated by $\varphi$. This closes to a holonomy symmetry generated by $\psi$, and consequently the relevant condition is of type I,

\begin{equation}
    P_4(\check R,F)_{\mu\nu[\rho|\sigma|}
    \varphi^\mu{}_{\kappa_1\kappa_2}
    \varphi^\nu{}_{\lambda_1\lambda_2]}=0~.
\end{equation}
This condition holds identically, as the skew-symmetrised product of the $\varphi$ tensors is $\mu$,$\nu$-symmetric. 

The mixed commutator of $\delta_\varphi$ with $\delta_\psi$, given in (\ref{commutator_phi_psi}), has a more involved structure. It contains a type II transformation generated by $S$, together with type III transformations associated with the pairs $(A,B)=(\varphi,C)$ and $I'=M'=K$. As shown in Section \ref{section_anomalies}, the type III contributions satisfy the type I condition. But due to the presence of a type II transformation, the consistency condition for the full commutator follows from (\ref{CC_type_II}) upon setting $L=\varphi$ and $M=\psi$, and yields
\begin{equation}\label{cc_phi_psi}
    2P_4(\check{R},F)_{\sigma[\rho\kappa_1\kappa_2}\varphi_{\lambda_1\lambda_2\lambda_3]} + P_4(\check{R},F)_{\mu\nu[\rho|\sigma|}\varphi^\mu{}_{\kappa_1\kappa_2}\psi^\nu{}_{\lambda_1\lambda_2\lambda_3]} = 0~.
\end{equation}
To verify this relation, we take the seven-dimensional Hodge dual over the six skew-symmetrised indices. Using $\psi=\star_7\varphi$ and the $G_2$ identities in Appendix \ref{subsection_identities}, the resulting expression vanishes identically. Hence, the consistency condition is satisfied without imposing any additional constraints on $P_4(\check R,F)$.

Finally, the commutator of two transformations generated by $\psi$ closes to a type II symmetry generated by $S$, a type IV symmetry generated by the Nijenhuis form $\mathcal N(\psi,\psi)$, and type III transformations associated with $(A,B)=(\psi,C)$ and $(I',M')=(\psi,K)$. As established in Section \ref{section_anomalies}, the type IV contribution satisfies the type I condition for the horizontal and $\hat{\nabla}$-covariantly constant Nijenhuis form considered here. Similarly, the type III contributions reduce to type I conditions. However, due to the presence of a type II symmetry, the corresponding condition follows from (\ref{CC_type_II}), with $L=M=\psi$, 
\begin{equation}\label{cc_psi_psi}
    6P_4(\check{R},F)_{\sigma[\rho\kappa_1\kappa_2}\psi_{\kappa_3\kappa_4\kappa_5\kappa_6]} + P_4(\check{R},F)_{\mu\nu[\rho|\sigma|}\psi^\mu{}_{\kappa_1\kappa_2\kappa_3}\psi^\nu{}_{\kappa_4\kappa_5\kappa_6]}=0~.
\end{equation}
Contracting with the Levi--Civita tensor and using the $G_2$ identities in Appendix \ref{subsection_identities}, one obtains
\begin{equation}\label{cc_psi_psi_2}
    P_4(\check{R},F)_{\mu\nu\rho\sigma}\varphi^{\mu\nu\rho}=0~.
\end{equation}
    
Since the left-hand side of (\ref{cc_psi_psi}) is totally skew-symmetric in seven transverse indices and the Hodge star is invertible, its contraction with the Levi--Civita tensor yields an equivalent condition. It therefore suffices to establish (\ref{cc_psi_psi_2}). The $G_2$-instanton conditions, $F_{\mu\nu}\varphi^{\mu\nu}{}_\rho=0$ and $\check R_{\mu\nu}\varphi^{\mu\nu}{}_\rho=0$, imply that the curvature 2-forms belong to $\Lambda^2_{\mathbf{14}}$, according to the decomposition $\Lambda^2=\Lambda^2_{\mathbf{7}}\oplus\Lambda^2_{\mathbf{14}}$ under $G_2$. As the wedge product of two 2-forms is symmetric, the 4-form $P_4(\check R,F)$ is determined by the symmetric tensor product $\mathrm{Sym}^2(\mathbf{14})=\mathbf{1}\oplus\mathbf{27}\oplus\mathbf{77}$. On the other hand, 4-forms decompose as $\Lambda^4=\Lambda^4_{\mathbf{1}}\oplus\Lambda^4_{\mathbf{7}}\oplus\Lambda^4_{\mathbf{27}}$. As the wedge product is $G_2$-equivariant, only representations common to both decompositions can contribute to $P_4(\check R,F)$, namely $\mathbf{1}$ and $\mathbf{27}$. The $\mathbf7$ component of $P_4(\check R,F)$ therefore vanishes, and since contraction with $\varphi$ selects this component, both (\ref{cc_psi_psi_2}) and the equivalent condition (\ref{cc_psi_psi}) are satisfied. 

This completes the analysis of the Wess--Zumino consistency conditions for the holonomy symmetry anomalies. Together with the cancellation mechanism discussed in Section \ref{section_anomalies}, these results establish that the anomalies are consistent and can be cancelled at one loop in the sigma model perturbation theory, under the assumption that the corrected Killing spinor bilinears satisfy the conditions (\ref{corrected_conditions}).

\section{Concluding remarks}\label{section_4}

We have presented the commutators of the transformations generated by $\hat{\nabla}$-covariantly constant forms that leave invariant heterotic sigma models on heterotic backgrounds with $G_2$ holonomy. We have shown that these transformations close to a W-algebra, as its structure constants are field-dependent, involving the conserved currents of the theory. Closure of the algebra requires additional symmetry generators that, unlike those of holonomy symmetries, are not generally constructed as Killing spinor bilinears. Compared with the corresponding algebras for the $SU$ holonomy backgrounds derived in \cite{Grimanellis:2023nav}, the algebra of the $G_2$ holonomy backgrounds contains an additional term associated with the Nijenhuis form. Remarkably, the Nijenhuis form is generally non-zero in the $SU$ case as well.

We derived the Wess--Zumino consistency conditions for the anomalies of the holonomy symmetries that arise due to the presence of chiral worldsheet fermions in the sigma model action and established the consistency of the anomalies at one loop. Assuming that the quantum-corrected Killing spinor bilinears satisfy the conditions (\ref{corrected_conditions}), the anomalies of the corresponding holonomy symmetries can be cancelled at the same loop order. This mechanism is compatible both with the anomaly cancellation of the gauge sector symmetries \cite{Hull:1985jv} and the preservation of the form of the Killing spinor equations of heterotic supergravity up to and including two loops in sigma model perturbation theory \cite{Bergshoeff:1989de}. Alternatively, for holonomy symmetries generated by Killing vectors, the associated anomalies can be cancelled by adding a finite local counterterm to the effective action, as in \cite{Grimanellis:2023nav}. The construction of analogous counterterms for the higher-degree holonomy symmetries in that work relies on the existence of complex structures, which are absent from the seven-dimensional base space of heterotic $G_2$ backgrounds. Consequently, these methods cannot be applied to cancel the anomalies of the symmetries generated by the 3-form and its dual. However, this does not necessarily rule out an alternative construction of appropriate counterterms that exploit the geometric properties of $G_2$ backgrounds.

Another distinction between holonomy symmetries generated by Killing vectors and those generated by higher-degree forms lies in their geometric interpretation. While the former, as generators of isometries, can be realised in the target space, the latter do not generally admit such a realisation. Instead, they act on the sigma model superfields through transformations involving their superspace derivatives. The Poisson bracket algebra of the associated conserved currents suggests that these higher-degree symmetries may admit a natural interpretation in the phase space of the theory \cite{Howe:1991ic}.

For completeness, we also comment on the case of trivial holonomy $\{1\}$. Such backgrounds admit ten $\hat\nabla$-parallel Killing vectors, whose associated sigma model transformations close under the commutator (\ref{commutator_KK}). Locally, the spacetime admits the structure of a ten-dimensional Lorentzian group manifold \cite{Gran:2005wf}. In the maximally supersymmetric case, $N=16$, where all parallel spinors are also Killing, the spacetime is locally isometric to Minkowski space \cite{Figueroa-OFarrill:2002ecq}, and these transformations generate an abelian algebra of translations. Furthermore, these symmetries remain non-anomalous upon quantisation.

Finally, at one loop, we have assumed that the anomalies of the holonomy symmetries (\ref{hol_anomaly}) do not receive additional contributions invariant under spacetime frame rotations and gauge transformations. As discussed in \cite{deLaOssa:2018iij,Grimanellis:2023nav}, such contributions may arise at higher orders in perturbation theory. It would be interesting to investigate whether they arise for the holonomy symmetries of $G_2$ backgrounds and, if so, whether they lead to modifications of the covariant constancy conditions satisfied by the corrected forms. Such an investigation would require a more detailed analysis of quantum corrections to the symmetry transformations, subject to the Wess–Zumino consistency conditions for the full anomaly.

\section*{Acknowledgements}
               
It is a pleasure to acknowledge the insightful suggestions of Georgios Papadopoulos, which proved particularly valuable in the course of this work. Special thanks are due to Athanasios Chatzistavrakidis for his trust and generous support, as well as for his constructive feedback. We are also grateful to Eric Bergshoeff and Alex Arvanitakis for their helpful comments. L.G. is supported by the Croatian Science Foundation (HRZZ) under the grant Higher Structures and Symmetries in Gauge and Gravity Theories (IP-2024-05-7921).

\appendix

\section{Notation and conventions}\label{appendix_A}

Given a $d$-dimensional manifold $M$, a $p$-form $\omega\in\Lambda^p(M)$ is defined as
\begin{equation}
    \omega = \frac{1}{p!}\omega_{\mu_1\dots\mu_p}e^{\mu_1}\wedge \dots \wedge e^{\mu_p}~,
\end{equation}
with a Hodge dual
\begin{equation}
    \star_d \omega = \frac{1}{p!(d-p)!}\omega_{\mu_1\dots\mu_p}\epsilon^{\mu_1\dots\mu_p}{}_{\mu_{p+1}\dots\mu_d} e^{\mu_{p+1}}\wedge \dots \wedge e^{\mu_d}~.
\end{equation}
The Levi--Civita tensor $\epsilon$ satisfies
\begin{equation}
    \epsilon^{\mu_1\dots \mu_q\nu_1\dots\nu_r}\epsilon_{\mu_1\dots \mu_q\rho_1\dots\rho_r} = q!(-1)^s\delta^{\nu_1\dots\nu_r}_{~\rho_1\dots \rho_r}~,
\end{equation}
where $d=q+r$, $s$ is the number of negative eigenvalues of the metric, and $\delta^{\nu_1\dots\nu_r}_{\rho_1\dots \rho_r}=r!\delta^{[\nu_1\dots\nu_r]}_{~\rho_1\dots \rho_r}$ the generalised Kronecker delta.

\subsection{\texorpdfstring{$G_2$}{G₂} identities}\label{subsection_identities}

The computation of the algebra of transformations generated by the $G_2$-invariant forms involves various contractions between them. Using the expressions (\ref{phi_psi_frame}), one obtains the following useful identities; see also \cite{Karigiannis2007FlowsOG}.
\paragraph{$\varphi-\varphi$ identities.}
\begin{align}
    \label{phi_phi_1}
    \varphi^{pij}\varphi_{pkl} &= 2\delta^i{}_{[k}\delta^j{}_{l]} - \psi^{ij}{}_{kl}~, \\
    \label{phi_phi_2}
    \varphi^{pqi} \varphi_{pqj} &= 6\delta^i{}_j ~, \\
    \varphi^{pqr}\varphi_{pqr} &= 42~.
\end{align} 
\paragraph{$\varphi-\psi$ identities.}
\begin{align}
    \varphi^{pij}\psi_{pklm} &= \frac{3}{2}\delta^{[i}{}_{[k}\varphi^{j]pq}\psi_{lm]pq}~, \label{phi_psi_1B} \\
    \varphi^{pij}\psi_{pklm} &= -6\delta^{[i}{}_{[k}\varphi^{j]}{}_{lm]}\label{phi_psi_1}~, \\
    \varphi^{pqi}\psi_{pqjk} &= -4\varphi^i{}_{jk} \label{phi_psi_2}~,\\
    \varphi^{pqr}\psi_{pqri} &= 0 \label{phi_psi_3}~.
\end{align}
\paragraph{$\psi-\psi$ identities.}
\begin{align}
    \psi^{pqij}\psi_{pqkl} &= 8\delta^i{}_{[k}\delta^j{}_{l]} - 2\psi^{ij}{}_{kl}\label{psi_psi_2}~, \\ 
    \quad \psi^{pqri}\psi_{pqrj} &= 24\delta^i{}_j\label{psi_psi_3}~, \\
    \psi^{pqrs} \psi_{pqrs} &= 168 ~.
\end{align}
The relation (\ref{phi_psi_1B}) follows from expressing $\varphi$ and $\psi$ as the Hodge duals of $\psi$ and $\varphi$, respectively, and can then be combined with (\ref{phi_psi_2}) to obtain (\ref{phi_psi_1}).

\subsection{Bianchi identities}

Our convention for the curvature tensor is 
\begin{equation}
    \hat{R}^\rho{}_{\sigma,\mu\nu} = \partial_\mu \hat{\Gamma}^\rho_{\nu\sigma} - \partial_\nu\hat{\Gamma}^\rho_{\mu\sigma} + \hat{\Gamma}^\rho_{\mu \lambda}\hat{\Gamma}^\lambda_{\nu\sigma} - \hat{\Gamma}^\rho_{\nu \lambda}\hat{\Gamma}^\lambda_{\mu\sigma}~, 
\end{equation}
with connection coefficients 
\begin{equation}
    \hat{\Gamma}^\rho_{\mu\nu} = \Gamma^\rho_{\mu\nu} + \frac{1}{2}H^\rho{}_{\mu\nu}~,
\end{equation}
and $\Gamma^{\rho}_{\mu\nu}$ the Levi--Civita connection coefficients. Moreover, the Wess--Zumino consistency conditions require the introduction of the curvature tensor $\check{R}$ for the connection
\begin{equation}
    \check{\Gamma}^\rho_{\mu\nu} = \Gamma^\rho_{\mu\nu} - \frac{1}{2}H^\rho{}_{\mu\nu}~.
\end{equation}
For completeness, we shall derive the Bianchi identities for both $R_+\equiv\hat{R}$ and $R_-\equiv\check{R}$, without assuming that $H$ is closed, as in (\ref{H_corrected}). Note that the curvature tensor is skew-symmetric in both pairs of indices 
\begin{equation}
    R_{\pm\rho\sigma,\mu\nu}=R_{\pm[\rho\sigma],[\mu\nu]}~,
\end{equation}
but the pair-exchange symmetry of the Riemann tensor no longer holds in general. Expressing $R_\pm$ in terms of the Riemann tensor, $H$ and its Levi--Civita covariant derivative as
\begin{equation}
    R^\rho_{\pm\sigma,\mu\nu} = R^\rho{}_{\sigma,\mu\nu} \pm \frac{1}{2}\left( \nabla_\mu H^\rho{}_{\nu\sigma} - \nabla_\nu H^\rho{}_{\mu\sigma} \right) + \frac{1}{4}\left( H^\rho{}_{\mu\lambda} H^\lambda{}_{\nu\sigma} - H^\rho{}_{\nu\lambda}H^\lambda{}_{\mu\sigma} \right)~,
\end{equation}
one finds
\begin{align}\label{R_ind_exch}
    R_{+\rho\sigma,\mu\nu}
    - R_{-\mu\nu,\rho\sigma}
    &= \frac{1}{2}\big(
    \nabla_\mu H_{\rho\nu\sigma}
    - \nabla_\nu H_{\rho\mu\sigma}
    + \nabla_\rho H_{\mu\sigma\nu}
    - \nabla_\sigma H_{\mu\rho\nu}
    \big) \nonumber\\
    &= -\frac{1}{2}dH_{\rho\sigma\mu\nu}~.
\end{align}

To derive the Bianchi identities,
one first evaluates the commutators of covariant derivatives acting on $X^\rho$ and $\nabla^\pm_\sigma X^\rho$
\begin{equation}
    [\nabla^\pm_\mu, \nabla^\pm_\nu]X^\rho = R^{\rho}_{\pm \kappa,\mu\nu} X^\kappa \mp H^\kappa{}_{\mu\nu} \nabla^\pm_\kappa X^\rho~,
\end{equation}
and
\begin{equation}
    [ \nabla^\pm_\mu, \nabla^\pm_\nu ]\nabla^\pm_\sigma X^\rho = - R^\kappa_{\pm\sigma,\mu\nu}\nabla^\pm_\kappa X^\rho + R^\rho_{\pm\kappa,\mu\nu}\nabla^\pm_\sigma X^\kappa \mp H^\lambda{}_{\mu\nu} \nabla^\pm_\lambda \nabla^\pm_\sigma X^\rho~.
\end{equation}
Then the Jacobi identity
\begin{equation}
    [\nabla^\pm_\mu,[\nabla^\pm_\nu,\nabla^\pm_\rho]] + [\nabla^\pm_\rho,[\nabla^\pm_\mu,\nabla^\pm_\nu]] + [\nabla^\pm_\nu,[\nabla^\pm_\rho,\nabla^\pm_\mu]] = 0 ~,
\end{equation}
gives
\begin{equation}
    \left( \nabla^\pm_{[\mu}R^\sigma_{\pm|\kappa|,\nu\rho]} \mp H^\lambda{}_{[\mu\nu} R^\sigma_{\pm |\kappa|,\rho]\lambda} \right) X^\kappa + \left( R^\lambda_{\pm[\mu,\nu\rho]} \mp \nabla^\pm_{[\mu}H^\lambda{}_{\nu\rho]} + H^\kappa{}_{[\mu\nu}H^\lambda{}_{\rho]\kappa}  \right) \nabla^\pm_\lambda X^\sigma = 0~.
\end{equation}
Since $X^\sigma$ is arbitrary, the two expressions in parentheses must vanish independently, yielding the two Bianchi identities for the torsionful connections $\nabla^\pm$
\begin{equation}
    R^\lambda_{\pm[\mu,\nu\rho]} =  \pm \nabla^\pm_{[\mu}H^\lambda{}_{\nu\rho]} - H^\kappa{}_{[\mu\nu}H^\lambda{}_{\rho]\kappa}~, \quad \nabla^\pm_{[\mu}R^\sigma_{\pm|\kappa|,\nu\rho]} = \pm H^\lambda{}_{[\mu\nu} R^\sigma_{\pm|\kappa|,\rho]\lambda} ~.
\end{equation}
Using (\ref{R_ind_exch}), the first Bianchi identity can be cast in the form 
\begin{equation}\label{Bianchi1}
    R_{\pm[\mu\nu,|\lambda|\rho]} = \mp\frac{1}{3}\nabla^\pm_\lambda H_{\mu\nu\rho} \mp \frac{1}{6}dH_{\lambda\mu\nu\rho}~.
\end{equation}

\bibliographystyle{JHEP}
\bibliography{sample}

@article{Grimanellis:2023nav,
    author = "Grimanellis, L. and Papadopoulos, G. and Perez-Bolanos, E.",
    title = "{W-symmetries, anomalies and heterotic backgrounds with $SU$ holonomy}",
    eprint = "2305.19793",
    archivePrefix = "arXiv",
    primaryClass = "hep-th",
    doi = "10.1016/j.nuclphysb.2023.116409",
    journal = "Nucl. Phys. B",
    volume = "998",
    pages = "116409",
    year = "2024"
}

@article{Papadopoulos:2023exe,
    author = "Papadopoulos, G. and Phillips, J.",
    title = "{W-symmetries, anomalies and heterotic backgrounds with non-compact holonomy}",
    eprint = "2311.18631",
    archivePrefix = "arXiv",
    primaryClass = "hep-th",
    doi = "10.1016/j.nuclphysb.2024.116618",
    journal = "Nucl. Phys. B",
    volume = "1005",
    pages = "116618",
    year = "2024"
}

@article{Howe:2010az,
    author = "Howe, P. S. and Papadopoulos, G. and Stojevic, V.",
    title = "{Covariantly constant forms on torsionful geometries from world-sheet and spacetime perspectives}",
    eprint = "1004.2824",
    archivePrefix = "arXiv",
    primaryClass = "hep-th",
    doi = "10.1007/JHEP09(2010)100",
    url = {https://link.springer.com/article/10.1007/JHEP09(2010)100},
    journal = "JHEP",
    volume = "09",
    pages = "100",
    year = "2010"
}

@article{Gran:2005wf,
    author = "Gran, U. and Lohrmann, P. and Papadopoulos, G.",
    title = "{The Spinorial geometry of supersymmetric heterotic string backgrounds}",
    eprint = "hep-th/0510176",
    archivePrefix = "arXiv",
    reportNumber = "KUL-TF-05-23",
    doi = "10.1088/1126-6708/2006/02/063",
    journal = "JHEP",
    volume = "02",
    pages = "063",
    year = "2006"
}

@article{Bergshoeff:1989de,
    author = "Bergshoeff, E. A. and de Roo, M.",
    title = "{The Quartic Effective Action of the Heterotic String and Supersymmetry}",
    reportNumber = "CERN-TH-5398-89",
    doi = "10.1016/0550-3213(89)90336-2",
    journal = "Nucl. Phys. B",
    volume = "328",
    pages = "439--468",
    year = "1989"
}

@article{Hull:1985jv,
    author = "Hull, C. M. and Witten, E.",
    editor = "Salam, A. and Sezgin, E.",
    title = "{Supersymmetric Sigma Models and the Heterotic String}",
    reportNumber = "PRINT-85-0617 (PRINCETON)",
    doi = "10.1016/0370-2693(85)90008-5",
    journal = "Phys. Lett. B",
    volume = "160",
    pages = "398--402",
    year = "1985"
}

@article{Gran:2018ijr,
    author = "Gran, U. and Gutowski, J. and Papadopoulos, G.",
    title = "{Classification, geometry and applications of supersymmetric backgrounds}",
    eprint = "1808.07879",
    archivePrefix = "arXiv",
    primaryClass = "hep-th",
    reportNumber = "DMUS-MP-18-06",
    doi = "10.1016/j.physrep.2018.11.005",
    journal = "Phys. Rept.",
    volume = "794",
    pages = "1--87",
    year = "2019"
}

@article{deLaOssa:2018iij,
    author = "de La Ossa, X. and Fiset, M. A.",
    title = "{$\mathcal{G}$-structure symmetries and anomalies in $(1,0)$ non-linear $\sigma$-models}",
    eprint = "1809.01138",
    archivePrefix = "arXiv",
    primaryClass = "hep-th",
    doi = "10.1007/JHEP01(2019)062",
    journal = "JHEP",
    volume = "01",
    pages = "062",
    year = "2019"
}

@article{Odake:1988bh,
    author = "Odake, Satoru",
    title = "{Extension of $N=2$ Superconformal Algebra and Calabi--Yau Compactification}",
    reportNumber = "UT-537-TOKYO",
    doi = "10.1142/S021773238900068X",
    journal = "Mod. Phys. Lett. A",
    volume = "4",
    pages = "557",
    year = "1989"
}

@article{Delius:1989nc,
    author = "Delius, G. W. and Rocek, M. and Sevrin, A. and van Nieuwenhuizen, P.",
    title = "{Supersymmetric sigma models with non-vanishing Nijenhuis tensor and their operator product expansion}",
    reportNumber = "ITP-SB-89-05",
    doi = "10.1016/0550-3213(89)90478-1",
    journal = "Nucl. Phys. B",
    volume = "324",
    pages = "523--531",
    year = "1989"
}

@article{Howe:1991im,
    author = "Howe, P. S. and Papadopoulos, G.",
    title = "{W-symmetries of a class of $d=2$, $N=1$ supersymmetric sigma models}",
    reportNumber = "QMW-PH-91-16",
    doi = "10.1016/0370-2693(91)90946-N",
    journal = "Phys. Lett. B",
    volume = "267",
    pages = "362--365",
    year = "1991"
}

@article{Howe:1991vs,
    author = "Howe, P. S. and Papadopoulos, G.",
    title = "{A Note on holonomy groups and sigma models}",
    reportNumber = "ITP-SB-91-11, QMW-PH-91-10",
    doi = "10.1016/0370-2693(91)90591-D",
    journal = "Phys. Lett. B",
    volume = "263",
    pages = "230--232",
    year = "1991"
}

@article{Howe:1991ic,
    author = "Howe, P. S. and Papadopoulos, G.",
    title = "{Holonomy groups and W-symmetries}",
    eprint = "hep-th/9202036",
    archivePrefix = "arXiv",
    reportNumber = "PRINT-92-0189 (KING'S-COLL.,LONDON)",
    doi = "10.1007/BF02097022",
    journal = "Commun. Math. Phys.",
    volume = "151",
    pages = "467--480",
    year = "1993"
}

@article{Gran:2007fu,
    author = "Gran, U. and Papadopoulos, G. and Roest, D. and Sloane, P.",
    title = "{Geometry of all supersymmetric type I backgrounds}",
    eprint = "hep-th/0703143",
    archivePrefix = "arXiv",
    doi = "10.1088/1126-6708/2007/08/074",
    journal = "JHEP",
    volume = "08",
    pages = "074",
    year = "2007"
}

@article{Moore:1984ws,
    author = "Moore, G. W. and Nelson, P. C.",
    title = "{The Aetiology of Sigma Model Anomalies}",
    reportNumber = "HUTP-84-A076",
    doi = "10.1007/BF01212688",
    journal = "Commun. Math. Phys.",
    volume = "100",
    pages = "83",
    year = "1985"
}

@article{Alvarez-Gaume:1985bbj,
    author = "Alvarez-Gaume, L. and Ginsparg, P. H.",
    title = "{Geometry anomalies}",
    reportNumber = "HUTP-85/A015",
    doi = "10.1016/0550-3213(85)90324-4",
    journal = "Nucl. Phys. B",
    volume = "262",
    pages = "439--462",
    year = "1985"
}

@article{Bagger:1985pw,
    author = "Bagger, J. and Nemeschansky, D. and Yankielowicz, S.",
    editor = "Salam, A. and Sezgin, E.",
    title = "{Anomaly constraints on nonlinear sigma models}",
    reportNumber = "SLAC-PUB-3588",
    doi = "10.1016/0550-3213(85)90326-8",
    journal = "Nucl. Phys. B",
    volume = "262",
    pages = "478",
    year = "1985"
}

@article{Wess:1971yu,
    author = "Wess, J. and Zumino, B.",
    title = "{Consequences of anomalous Ward identities}",
    doi = "10.1016/0370-2693(71)90582-X",
    journal = "Phys. Lett. B",
    volume = "37",
    pages = "95--97",
    year = "1971"
}

@article{Green:1984sg,
    author = "Green, M. B. and Schwarz, J. H.",
    title = "{Anomaly Cancellation in Supersymmetric $D=10$ Gauge Theory and Superstring Theory}",
    reportNumber = "CALT-68-1182",
    doi = "10.1016/0370-2693(84)91565-X",
    journal = "Phys. Lett. B",
    volume = "149",
    pages = "117--122",
    year = "1984"
}

@article{Howe:1987nw,
    author = "Howe, P. S. and Papadopoulos, G.",
    title = "{Anomalies in two-dimensional supersymmetric non-linear $\sigma$ models}",
    reportNumber = "PRINT-87-0620 (KING'S-COLL)",
    doi = "10.1088/0264-9381/4/6/027",
    journal = "Class. Quant. Grav.",
    volume = "4",
    pages = "1749--1766",
    year = "1987"
}

@article{Alvarez-Gaume:1983ict,
    author = "Alvarez-Gaume, L. and Ginsparg, P. H.",
    title = "{The topological meaning of non-abelian anomalies}",
    reportNumber = "HUTP-83/A081",
    doi = "10.1016/0550-3213(84)90487-5",
    journal = "Nucl. Phys. B",
    volume = "243",
    pages = "449--474",
    year = "1984"
}

@article{Atiyah:1984tf,
    author = "Atiyah, M. F. and Singer, I. M.",
    title = "{Dirac Operators Coupled to Vector Potentials}",
    doi = "10.1073/pnas.81.8.2597",
    journal = "Proc. Natl. Acad. Sci. U.S.A.",
    volume = "81",
    pages = "2597--2600",
    year = "1984"
}

@article{Adler:1969gk,
    author = "Adler, S. L.",
    title = "{Axial-Vector Vertex in Spinor Electrodynamics}",
    doi = "10.1103/PhysRev.177.2426",
    journal = "Phys. Rev.",
    volume = "177",
    pages = "2426--2438",
    year = "1969"
}

@article{Bell:1969ts,
    author = "Bell, J. S. and Jackiw, R.",
    title = "{A PCAC puzzle: $\pi^0 \to \gamma \gamma$ in the $\sigma$-model}",
    doi = "10.1007/BF02823296",
    journal = "Nuovo Cim. A",
    volume = "60",
    pages = "47--61",
    year = "1969"
}

@article{Bardeen:1969md,
    author = "Bardeen, W. A.",
    title = "{Anomalous Ward Identities in Spinor Field Theories}",
    doi = "10.1103/PhysRev.184.1848",
    journal = "Phys. Rev.",
    volume = "184",
    pages = "1848--1857",
    year = "1969"
}

@article{Alvarez-Gaume:1981exa,
    author = "Alvarez-Gaume, L. and Freedman, D. Z. and Mukhi, S.",
    title = "{The Background Field Method and the Ultraviolet Structure of the Supersymmetric Nonlinear $\sigma$-Model}",
    reportNumber = "Print-81-0045 (MIT)",
    doi = "10.1016/0003-4916(81)90006-3",
    journal = "Annals Phys.",
    volume = "134",
    pages = "85",
    year = "1981"
}

@article{Sakamoto:1984zk,
    author = "Sakamoto, M.",
    title = "{$N=1/2$ supersymmetry in two dimensions}",
    reportNumber = "KYUSHU-84-HE-3",
    doi = "10.1016/0370-2693(85)91396-6",
    journal = "Phys. Lett. B",
    volume = "151",
    pages = "115--118",
    year = "1985"
}

@article{Siegel:1983es,
    author = "Siegel, W.",
    title = "{Manifest Lorentz Invariance Sometimes Requires Nonlinearity}",
    reportNumber = "UCB-PTH-83/22",
    doi = "10.1016/0550-3213(84)90453-X",
    journal = "Nucl. Phys. B",
    volume = "238",
    pages = "307--316",
    year = "1984"
}

@article{Frolicher:1956nk1,
author = "Frölicher, A. and Nijenhuis, A.",
title = "{Theory of Vector-Valued Differential Forms: Part I. Derivations in the Graded Ring of Differential Forms}",
journal = "Indag. Math. (Proc.)",
volume = "59",
pages = "338-359",
year = "1956",
issn = "1385-7258",
doi = "https://doi.org/10.1016/S1385-7258(56)50046-7",
url = "https://www.sciencedirect.com/science/article/pii/S1385725856500467"
}

@article{Nijenhuis:1955dl,
title = "{Jacobi-type identities for bilinear differential concomitants of certain tensor fields. I}",
journal = "Indag. Math. (Proc.)",
volume = {58},
pages = {390-397},
year = {1955},
issn = {1385-7258},
doi = {https://doi.org/10.1016/S1385-7258(55)50054-0},
url = {https://www.sciencedirect.com/science/article/pii/S1385725855500540},
author = {Nijenhuis, A.}
}

@article{Nijenhuis:1955dm,
title = "{Jacobi-type identities for bilinear differential concomitants of certain tensor fields. II}",
journal = "Indag. Math. (Proc.)",
volume = {58},
pages = {398-403},
year = {1955},
issn = {1385-7258},
doi = {https://doi.org/10.1016/S1385-7258(55)50055-2},
url = {https://www.sciencedirect.com/science/article/pii/S1385725855500552},
author = {Nijenhuis, A}
}

@article{Hull:1986xn,
    author = "Hull, C. M. and Townsend, P. K.",
    title = "{World-Sheet Supersymmetry and Anomaly Cancellation in the Heterotic String}",
    reportNumber = "PRINT-86-1240 (CAMBRIDGE)",
    doi = "10.1016/0370-2693(86)91493-0",
    journal = "Phys. Lett. B",
    volume = "178",
    pages = "187--192",
    year = "1986"
}

@article{Honerkamp:1971sh,
    author = "Honerkamp, J.",
    title = "{Chiral multiloops}",
    doi = "10.1016/0550-3213(72)90299-4",
    journal = "Nucl. Phys. B",
    volume = "36",
    pages = "130--140",
    year = "1972"
}

@article{Friedan:1980jm,
    author = "Friedan, D. H.",
    title = "{Nonlinear Models in $2+\varepsilon$ Dimensions}",
    reportNumber = "LBL-11517, UMI-81-13038",
    doi = "10.1016/0003-4916(85)90384-7",
    journal = "Annals Phys.",
    volume = "163",
    pages = "318",
    year = "1985"
}

@article{Hull:1993ct,
    author = "Hull, C. M. and Papadopoulos, G. and Townsend, P. K.",
    title = "{Potentials for $(p,0)$ and $(1,1)$ supersymmetric sigma models with torsion}",
    eprint = "hep-th/9307013",
    archivePrefix = "arXiv",
    reportNumber = "QMW-93-11, DAMTP-R-93-8, KCL-93-5",
    doi = "10.1016/0370-2693(93)90327-E",
    journal = "Phys. Lett. B",
    volume = "316",
    pages = "291--297",
    year = "1993"
}

@article{Howe:2006si,
    author = "Howe, P. S. and Stojevic, V.",
    title = "{On the symmetries of special holonomy sigma models}",
    eprint = "hep-th/0606270",
    archivePrefix = "arXiv",
    reportNumber = "KCL-TH-06-05",
    doi = "10.1088/1126-6708/2006/12/045",
    journal = "JHEP",
    volume = "12",
    pages = "045",
    year = "2006"
}

@incollection{Zumino1984,
  author    = {Zumino, B.},
  title     = {Chiral Anomalies and Differential Geometry},
  booktitle = {Relativity, Groups and Topology II},
  editor    = {B. S. DeWitt and R. Stora},
  publisher = {North-Holland},
  address   = {Amsterdam},
  year      = {1984},
  pages     = {1291--1322},
}

@article{Sen:1986mg,
    author = "Sen, A.",
    title = "{$(2, 0)$ Supersymmetry and Space-Time Supersymmetry in the Heterotic String Theory}",
    reportNumber = "SLAC-PUB-3919",
    doi = "10.1016/0550-3213(86)90214-2",
    journal = "Nucl. Phys. B",
    volume = "278",
    pages = "289--308",
    year = "1986"
}

@article{Karigiannis2007FlowsOG,
  author = "Karigiannis, S.",
  title = "{Flows of $G_2$-Structures, I}",
  doi = "10.1093/qmath/han020",
  journal = "Q. J. Math.",
  volume = "60",
  pages = "487-522",
  year = "2007",
  url={https://api.semanticscholar.org/CorpusID:119689206}
}

@article{Friedrich:2001nh,
    author = "Friedrich, T. and Ivanov, S.",
    title = "{Parallel spinors and connections with skew-symmetric torsion in string theory}",
    eprint = "math/0102142",
    archivePrefix = "arXiv",
    reportNumber = "SFB-288-492",
    doi = "https://dx.doi.org/10.4310/AJM.2002.v6.n2.a5",
    journal = "Asian J. Math.",
    volume = "6",
    pages = "303--336",
    year = "2002"
}

@article{Friedrich:2001yp,
    author = "Friedrich, T. and Ivanov, S.",
    title = "{Killing spinor equations in dimension 7 and geometry of integrable $G_2$-manifolds}",
    eprint = "math/0112201",
    archivePrefix = "arXiv",
    doi = "10.1016/S0393-0440(03)00005-6",
    journal = "J. Geom. Phys.",
    volume = "48",
    pages = "1",
    year = "2003"
}

@article{BARDEEN1984421,
title = {Consistent and covariant anomalies in gauge and gravitational theories},
journal = {Nucl. Phys. B},
volume = {244},
number = {2},
pages = {421-453},
year = {1984},
issn = {0550-3213},
doi = {https://doi.org/10.1016/0550-3213(84)90322-5},
url = {https://www.sciencedirect.com/science/article/pii/0550321384903225},
author = {Bardeen, W. A. and Zumino, B.}
}

@article{Shatashvili:1994zw,
    author = "Shatashvili, S. L. and Vafa, C.",
    title = "{Superstrings and manifold of exceptional holonomy}",
    eprint = "hep-th/9407025",
    archivePrefix = "arXiv",
    reportNumber = "HUTP-94-A016, IASSNS-HEP-94-47",
    doi = "10.1007/BF01671569",
    journal = "Selecta Math.",
    volume = "1",
    pages = "347",
    year = "1995"
}

@article{Figueroa-OFarrill:1996tnk,
    author = "Figueroa-O'Farrill, J. M.",
    title = "{A Note on the extended superconformal algebras associated with manifolds of exceptional holonomy}",
    eprint = "hep-th/9609113",
    archivePrefix = "arXiv",
    reportNumber = "QMW-PH-96-25",
    doi = "10.1016/S0370-2693(96)01506-7",
    journal = "Phys. Lett. B",
    volume = "392",
    pages = "77--84",
    year = "1997"
}

@article{Fiset:2021azq,
    author = "Fiset, M. A. and Gaberdiel, M. R.",
    title = "{Deformed Shatashvili--Vafa algebra for superstrings on AdS$_3\times{\cal M}_7$}",
    eprint = "2101.10327",
    archivePrefix = "arXiv",
    primaryClass = "hep-th",
    doi = "10.1007/JHEP05(2021)156",
    journal = "JHEP",
    volume = "05",
    pages = "156",
    year = "2021"
}

@article{delaOssa:2024cgo,
    author = "de la Ossa, X. and Galdeano, M. and Marchetto, E.",
    title = "{$ \mathcal{SW} $-algebras and strings with torsion}",
    eprint = "2412.13904",
    archivePrefix = "arXiv",
    primaryClass = "hep-th",
    reportNumber = "DESY-24-197",
    doi = "10.1007/JHEP04(2025)157",
    journal = "JHEP",
    volume = "04",
    pages = "157",
    year = "2025"
}

@article{Figueroa-OFarrill:2002ecq,
    author = "Figueroa-O'Farrill, J. M. and Papadopoulos, G.",
    title = "{Maximally supersymmetric solutions of ten-dimensional and eleven-dimensional supergravities}",
    eprint = "hep-th/0211089",
    archivePrefix = "arXiv",
    reportNumber = "EMPG-02-16",
    doi = "10.1088/1126-6708/2003/03/048",
    journal = "JHEP",
    volume = "03",
    pages = "048",
    year = "2003"
}

@article{Gillard:2003jh,
    author = "Gillard, J. and Papadopoulos, G. and Tsimpis, D.",
    title = "{Anomaly, fluxes and $(2,0)$ heterotic string compactifications}",
    eprint = "hep-th/0304126",
    archivePrefix = "arXiv",
    doi = "10.1088/1126-6708/2003/06/035",
    journal = "JHEP",
    volume = "06",
    pages = "035",
    year = "2003"
}

@article{Alvarez-Gaume:1984zlq,
    author = "Alvarez-Gaume, Luis and Ginsparg, Paul H.",
    editor = "Salam, A. and Sezgin, E.",
    title = "{The Structure of Gauge and Gravitational Anomalies}",
    reportNumber = "HUTP-84/A016",
    doi = "10.1016/0003-4916(85)90087-9",
    journal = "Annals Phys.",
    volume = "161",
    pages = "423",
    year = "1985",
    note = "[Erratum: Annals Phys. 171, 233 (1986)]"
}

\end{document}